\documentclass[sigconf]{acmart}
\usepackage{multirow}
\usepackage[nameinlink]{cleveref}
\usepackage{fancybox}
\AtBeginDocument{%
  \providecommand\BibTeX{{%
    \normalfont B\kern-0.5em{\scshape i\kern-0.25em b}\kern-0.8em\TeX}}}

\definecolor{myBlue}{RGB}{20, 96, 178}
\acmConference[UIST '24]{The ACM Symposium on User Interface Software and Technology}{Oct 13--16,
  2024}{Pittsburgh, PA}

\newcommand{\rev}[1]{{\color{black}#1}} %\color{blue}

\newcommand{\figref}[1]{Fig.~\ref{fig:#1}}

\makeatletter
\renewcommand{\paragraph}{%
  \@startsection{paragraph}{4}%
  {\z@}{0.60ex \@plus 1ex \@minus .15ex}{-1em}%
  {\normalfont\normalsize\bfseries}%
}
\makeatother

\begin{document}

\title{SonifyAR: Context-Aware Sound Generation in Augmented Reality}

\renewcommand{\shortauthors}{Su et al.}
\author{Xia Su}
\authornote{Major work completed during an internship at Adobe Research.}
\affiliation{%
  \institution{University of Washington}
  \city{Seattle}
  \state{Washington}
  \country{USA}
}
\email{xiasu@cs.washington.edu}
\author{Jon E. Froehlich}
\affiliation{%
  \institution{University of Washington}
  \city{Seattle}
  \state{Washington}
  \country{USA}
}
\email{jonf@cs.washington.edu}

\author{Eunyee Koh}
\affiliation{%
  \institution{Adobe Research}
  \city{San Jose}
  \state{California}
  \country{USA}
}
\email{eunyee@adobe.com}
\author{Chang Xiao}
\affiliation{%
  \institution{Adobe Research}
  \city{San Jose}
  \state{California}
  \country{USA}
}
\email{cxiao@adobe.com}

\begin{abstract}

Sound plays a crucial role in enhancing user experience and immersiveness in Augmented Reality (AR). However, current platforms lack support for AR sound authoring due to limited interaction types, challenges in collecting and specifying context information, and difficulty in acquiring matching sound assets. We present SonifyAR, an LLM-based AR sound authoring system that generates context-aware sound effects for AR experiences. SonifyAR expands the current design space of AR sound and implements a \textit{Programming by Demonstration} (PbD) pipeline to automatically collect contextual information of AR events, including virtual-content-semantics and real-world context. This context information is then processed by a large language model to acquire sound effects with \textit{Recommendation}, \textit{Retrieval}, \textit{Generation}, and \textit{Transfer} methods. To evaluate the usability and performance of our system, we conducted a user study with eight participants and created five example applications, including an AR-based science experiment, and an assistive application for low-vision AR users.
\end{abstract}




\begin{teaserfigure}
  \includegraphics[width=\textwidth]{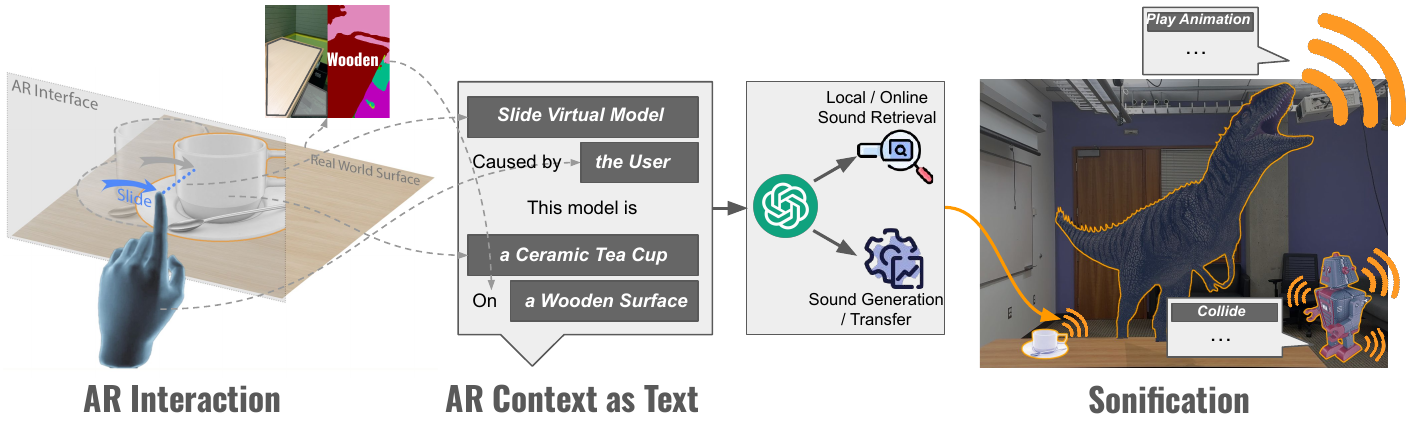}
  \caption{\rev{SonifyAR is a custom AR sound authoring pipeline that generates context-matching sounds for AR events \textit{in situ} using generative AI. For example, imagine sliding an AR tea cup across a real-world surface such as a wood table. SonifyAR observes this user action (a slide gesture), the action source (the user), and the action target (a virtual ceramic teacup), infers scene information such as the surface material (a wood table), and uses a custom AI backend to \textit{recommend}, \textit{retrieve}, \textit{generate}, or \textit{sound-style transfer} sound effects.}}
  \Description{SonifyAR pipeline overview. From left to right: an AR sliding operation on a virtual ceramic cup on a real-world table; A text box containing the context information of this interaction; SonifyAR use retrieve-based and generation-based methods to generate sound; The sound can be tested and experienced in AR.}
  \label{fig:teaser}
\end{teaserfigure}

% Xia's old text
% With SonifyAR, the user firsts interact with the AR Scene, while SonifyAR identifies potential sound-producing events and collects context information as text. This context text, including the real-world environment information like surface material, is then processed by LLM, which guides the sound acquisition process using multiple models and methods. The acquired sounds are then used to sonify AR events.

\maketitle

\section{Introduction}
%In recent years, Augmented Reality (AR) has evolved into an interdisciplinary domain, offering users multi-modal information to enhance immersive experiences. Beyond the visual component, auditory feedback has been demonstrated to play a pivotal role in heightening immersion [CITE]. 

%Sound, as a effective addition to human-computer interfaces \cite{zhou_role_2007}, has been proved powerful in improving Augmented Reality (AR) experiences [cites]. 
%The role of sound in Augmented Reality (AR) has grown profoundly, becoming a pivotal element that adds depth and immersion to user experiences,
Sound is a critical but often overlooked element in Augmented Reality (AR). AR-based sound can improve immersion and overall user experience~\cite{roginska2017immersive} and support depth perception and task completion \cite{zhou_role_2007}, search and navigation~\cite{ruminski2015experimental}, and even assistance for people with low vision ~\cite{ribeiro_auditory_2012}. Despite its importance, current AR authoring platforms like \textit{Reality Composer} \cite{realitycomposer}, \textit{Adobe Aero} \cite{adobeaero}, and \textit{Unity Mars} \cite{UnityMARS2023} provide only rudimentary sound support. Specifically, we identified three critical gaps with existing AR authoring tools:

\begin{enumerate}
    \item \textbf{Limited real-world context.} Existing systems typically support action triggers linked to virtual objects in AR but lack support for real-world contextual information (\textit{e.g.}, a virtual toy robot traversing diverse indoor surfaces like wood, carpet, or glass.). 
    \item \textbf{Limited interaction specification.} Existing systems provide only pre-defined interaction triggers like \textit{``Tap''} and \textit{``Proximity Enter''}. This limits a creator's ability to specify interactions outside the provided options, especially those that involve environmental context (\textit{e.g.,} the user ``slides'' virtual chalk on a real-world blackboard). 
    \item  \textbf{Limited sound sources.} Existing systems are limited by the sound assets available in their libraries and the scarcity of suitable sound resources online. Thus, AR authors struggle to find appropriate sounds for distinct AR events (\textit{e.g.,} reproducing the wing flutter of a virtual dragonfly or simulating the eating sound of a virtual dinosaur). 
\end{enumerate}

To address these challenges, we present \textit{SonifyAR}: a context-aware AR sound authoring system using \textit{GPT-4} \cite{openai2024gpt4} and a \textit{text2audio} diffusion model called \textit{AudioLDM} \cite{liu2023audioldm}. SonifyAR makes the following key technical advancements:

%\textbf{Expansion of AR sound interaction space.} Inspired by Jain et al. \cite{jain_taxonomy_2021}, we analyze the auditory design space of AR interactions and describe the interaction space as a triad between user, virtuality and reality. This triad greatly expands the existing AR sound interaction spaces.

\textbf{First, context collection using PbD.} SonifyAR adopts a \textit{Programming by Demonstration} (PbD) \cite{10.1145/291080.291104} pipeline to simplify the specification of complex AR interactions. The PbD pipeline enables users to demonstrate AR sound interactions while the system automatically detects the action and collects context information. For example, if a creator wants to sonify the stomping of a walking robot, they can position the virtual robot on the target (physical) surface. As the robot walks, the collision between robot's feet and the surface, as well as the context information like the robot's attributes and the surface's material, is captured by SonifyAR .

\textbf{Second, context as text.} To utilize AR context information stemming from multiple sources (\textit{e.g.} user action, virtual object, real-world environment) and in different formats (\textit{e.g.} categorical, 3D shape, image), we use text as the universal medium to encompass all context information. For example, we generate an LLM prompt using the following template: ``\textit{This event is [Event Type], caused by [Source]. This event casts on [Target Object]. [Additional Information on Involved Entities]}''.

\textbf{Finally, LLM-based sound acquisition.} We integrate a suite of four sound acquisition methods: \textit{recommend}, \textit{retrieve}, \textit{generate}, and \textit{transfer}, all controlled by the underlying LLM. For each AR sound interaction, the context information as text is fed to the LLM for processing. The LLM then provides text prompts that control the suite of four sound acquisition methods, which automatically provide matching sound assets for the AR interaction.

To author AR sound with SonifyAR, the creator initiates AR interactions and the system automatically lists sound options based on context. For example, imagine trying to make a virtual tea cup chime accordingly when being slid on a table (\autoref{fig:teaser}). In traditional approaches, the creator would need to manually specify this sliding action and find a sound asset to match the chiming ceramic cup. But with SonifyAR, the creator could demonstrate this sliding action and the chime sounds are generated automatically for the creator to choose from. 

To explore the potential of SonifyAR, we conducted two evaluations: first, a qualitative user study with eight designers to examine the usability of and reactions to the SonifyAR prototype; second, we apply SonifyAR across five key user scenarios: from AR education to improving AR headset safety. Our user study highlights the potential of SonifyAR and the generated sounds while identifying key areas of improvement such as sound quality and interface design. Our application scenarios demonstrate the breadth and potential of SonifyAR incorporating automatically acquired sounds seamlessly into various AR experiences—a practice that would otherwise take significant manual time and effort.

%The contributions of this paper are summarized as follows:
%First, we explored the landscape of sound authoring in AR, identifying a noticeable gap between the existing authoring tools and the broader potential opportunities in the design space. Second, to address this gap, we introduce an AR sound authoring pipeline that incorporate context information of the environment, virtual object, and user action. Lastly, we implement this proposed pipeline as an LLM and generative model based sound authoring system. This implementation is evaluated with eight participants for usability and future application opportunities.  

Our work makes three primary contributions: (1) a design space for AR sound authoring tools, which highlights existing gaps in the literature; (2) a novel context-aware AR sound authoring pipeline using generative AI, called SonifyAR, that incorporates sensed environmental cues like surface material, virtual object semantics, and user action; (3) findings from a user study with eight designers and five application examples that demonstrate SonifyAR's potential and key advancements in this space. To our knowledge, we are the first system to offer automatic, in-context sound authoring for AR. 

\section{Related Work}
We cover prior work in AR sound authoring, AI-based sound generation techniques, and context-awareness in AR. 
\subsection{AR Sound Authoring}
Sound in AR provides many benefits from directing the user's attention and enhancing immersion~\cite{roginska2017immersive,zhou_role_2007}  to creating interactive time-varying experiences and increasing accessibility \cite{serafin_sonic_2018,ruminski2015experimental}. However, as mentioned, the current landscape of AR authoring tools reveals deficiencies in how sound is supported.

%These authoring tools can be split into coded and non-coded. Coded, like Unity and Unreal, can provide basically any support for AR sound but. The non-coded, which are built for regular users, provides only selected common conditions thus fail to cover many cases. We focus on the scope of non-coded tools in this work.
Current tools highlight a variety of AR sound authoring methods.  \cite{monteiro2023teachable,nebeling_trouble_2018}. Some---like \textit{Adobe Aero} \cite{adobeaero} and \textit{Apple Reality Composer} \cite{realitycomposer}---are easy to use, require no coding skills, and can only attach user-provided sound with specific AR event triggers \cite{haloar,arvid,adobeaero,realitycomposer}. Others---like \textit{Unity} \cite{unity-website} and \textit{Unreal Engine} \cite{unreal-engine-website}---are heavily code-based and require significant technical skills, but provide substantial flexibility in designing AR sound interactions, including the ability to set parameters like sound decay and code-specific conditions for activating sound assets. One common unifying approach \textit{Programming by Specification} (PbS) methodology \cite{monteiro2023teachable}, where creators define AR sound content and play conditions during the design phase. %, which is also addressed as the ``\textit{trigger-action mechanism}'' \cite{monteiro2023teachable}. A ``\textit{trigger}'', such as tapping an AR object or a specific location in the scene, sets the condition for launching certain actions. Subsequently, an ``action'', whether it be the playing of an animation or a sound effect, responds to the earlier trigger. 
This allows creators to precisely define the conditions for triggers and the consequent sound effects.

%Problem! Restate the three points but with more details.
As mentioned in introduction, however, critical gaps exist. These gaps are partially due to the limitation of the PbS methodology, since the triggers specified in the design stage naturally lack richness in real-world context, which is crucial for AR sound realism. Also, the trade-offs between supported trigger types and the tool's usability make most tools lack coverage for many sound-producing AR events. Additionally, all sound assets need to be manually selected or created, tasking creators with sourcing suitable sound effects to match the specific triggers. The SonifyAR system aims to address these gaps.

\subsection{Sound Acquisition}
%new version here
Although no work has explored sound generation with AR events as input, sound generation conditioned from other input modalities like images, 3D models, text, and videos has been widely explored. Various cross-modal generative models like RNN \cite{Zhou_2018_CVPR}, GAN \cite{engel2019gansynth,kumar2019melgan,ghose2022foleygan}, VAE \cite{dhariwal2020jukebox}, and most recently diffusion models \cite{kong2020diffwave,liu2023audioldm,yang2023diffsound}, are utilized to generate sound. Despite different model architectures, one common training goal is the mapping between latent representations of input modalities and output sound. Thus, sounds can be generated from prompts like images \cite{10096023}, videos \cite{Zhou_2018_CVPR,ghose2022foleygan}, and text \cite{liu2023audioldm,yang2023diffsound}. For example, \textit{FoleyGAN} \cite{ghose2022foleygan} conditions action sequences of input videos to generate visually aligned, realistic soundtracks. \textit{AudioLDM} \cite{liu2023audioldm} utilizes the CLAP-based \cite{wu2023large} latent space to embed input text and generate matching sound with a diffusion model. Unlike cross-modal generative methods, another promising approach is physics-based sound synthesis \cite{raghuvanshi_interactive_2006,roodaki_sonifeye_2017,ren_example-guided_2013,jin_deep-modal_2020,jin_neuralsound_2022,liu_sound_2021,diaz_rigid-body_2022}. For example, the spring-mass model \cite{raghuvanshi_interactive_2006}---recently improved with deep learning \cite{jin_deep-modal_2020}---can provide realistic sound simulation given a 3D model and material properties. 

Besides synthesis, retrieval methods can also acquire matching sound for input conditions. For example, Koepke \textit{et al.}~\cite{koepke2022audio} trained cross-modal embedding models to support text-to-audio retrieval. Multi-modal models like VAST ~\cite{chen2023vast} also support text-to-audio retrieval that could potentially be used to provide sound assets for AR experiences. Recent work, called \textit{Soundify} \cite{lin2023soundify}, utilizes retrieved sound assets to sonify videos. The authors claim that although sound retrieval does not ensure coverage for all input descriptions, the retrieved sound assets usually surpass the synthesis results in terms of quality. In this case, combining generation and retrieval in the sound acquisition pipeline seems to be a natural method that integrates the merits of both. We explore this possibility with SonifyAR.

Unlike other existing sound acquisition methods, AR poses unique challenges in the matching of input condition and output sound. The multi-modal nature of AR interactions, including the information of the virtual object (\textit{e.g.} a toy robot with a walking animation), real environment (\textit{e.g.} a living room with a wooden table), and user action (\textit{e.g.} user taps on a virtual model), needs to be carefully processed into the sound acquisition pipeline to result in matching sound assets. 

\subsection{Context-awareness in AR}

In AR, context-awareness is an important property that allows AR content to be dynamically adjusted according to context information, such as location, scene semantics, and human factors. Context-awareness aims to adapt an application not only to the user but also to their environment and specific needs with the goal of improved usability and immersive \cite{krings_development_2020}. Previous research has explored various aspects of context-awareness in AR, highlighting its importance across different applications.
%The target is to show how existing work address context awareness, while neglecting the sound modality
%[Emphasize there's no dedicated tool for sound authoring in AR]
% Acquiring context-matching sound would require high level of context-awareness in AR.
%, which refers to applications adjusting their functionalities to align with the current context, enhancing usability and tailoring experiences for different operating conditions~\cite{abowd1999towards}.
% In AR, context-awareness ensures that the application adapts not only to the user but also to their environment and specific needs, making the application more user-friendly and immersive \cite{krings_development_2020}. There is extensive prior work on AR-based context awareness, including location, scene semantics and human factors. 

For example, the popular AR game \textit{Pokem\'{o}n Go}\cite{pokemon_go} shows game contents based on user's physical location. A more common practice in AR research is adapting virtual contents to scene semantics---essentially understanding the objects and their locations in the user's environment. Qian \textit{et al.} \cite{qian2022scalar} introduced an authoring system that helps user create adaptive AR experiences to surrounding scene semantics. Lang \textit{et al.} \cite{lang2019virtual} analyzes scene semantics to guide the positioning of virtual agents. Similarly, Liang \textit{et al.}~\cite{liang2021scene} places virtual pets into AR based on scene geometry and semantics. This realm of similar work \cite{han2020live,tahara_retargetable_2020} usually captures and analyzes the surrounding environment and designs algorithms that determines possible interaction between virtual content and real-world scene. Lindbauer \textit{et al.} \cite{lindlbauer2019context} push the AR context-awareness further by including task and cognitive load as context to adjust level of detail of AR virtual interface.

Although there has been substantial discussion in the realm of AR context-awareness, we observed that prior work generally focuses on the visual modality of AR, neglecting sound. Additionally, existing work primarily concentrates on adjusting and arranging virtual content rather than generating new content. We aim to fill this gap by designing a sound generation pipeline that processes context information to produce matching sound assets.

\section{The Design of SonifyAR}
To design an effective sound authoring system that generates matching sounds for AR interaction, three main research questions need to be answered. 
\begin{itemize}
    \item What are the target AR interactions that should produce sound effects? 
     \item How can we efficiently acquire sound assets for these AR interactions? 
    \item How can these AR interactions be specified and represented in the authoring process?

\end{itemize}
We aim to elaborate on these questions and the respective design rationale in this section.

\subsection{The Triad of User, Virtuality, and Reality}
We began by examining the design space of AR interaction sounds. Drawing on Jain \textit{et al.'s} \cite{jain_taxonomy_2021} sound taxonomy in VR, we attempted a similar analysis for AR sound interaction. We first investigated several widely used code-free AR authoring tools for their AR sound capability, including \textit{Adobe Aero} \cite{adobeaero}, \textit{Apple Reality Composer} \cite{realitycomposer}, and \textit{Halo AR} \cite{haloar}. We found that these tools support AR sound related to either \textit{virtual content interaction} or \textit{AR-based spatial anchors} (\textit{e.g.,}play a sound when a user gets close to a certain position, scans a QR code, or taps on a virtual model). 

The AR sound research literature \cite{dam2024taxonomy,rakkolainen2021technologies,filus2012using}, however, envisions a much broader scope. Rakkolainen \textit{et al.}~\cite{rakkolainen2021technologies} classified AR audio into five classes: \textit{ambient}, \textit{directional}, \textit{musical}, \textit{speech}, and \textit{noise}. 
Dam \textit{et al.'s} \cite{dam2024taxonomy} taxonomy of \textit{Audio Augmented Reality} (AAR), defines AR-based sound across  \textit{Environment Connected} (sound maintaining awareness and interaction with physical environment), \textit{Goal Directed} (sound assisting users' primary goal), and \textit{Context Adapted} (sound adaptive to users' immediate reality), as well as their intersection. This taxonomy emphasizes that AR sound is often rooted in contextual information from the real world environment and user intention. 

Drawing on this prior work, we introduce a novel AR sound design space that helps emphasize the sonic interplay between three key factors of AR experiences: \textit{reality}, denoting the physical environment in the real world that hosts the AR experience; \textit{virtuality}, referring to the virtual object(s) placed in the AR experience; and \textit{the user}, representing the individual interacting with the AR experience. Below, we expand on each element of this proposed design space and indicate current levels of support in the existing code-free AR sound authoring tools~\cite{adobeaero, realitycomposer,haloar}.

\begin{figure}[t]
    \centering
    \includegraphics[width=0.5\textwidth]{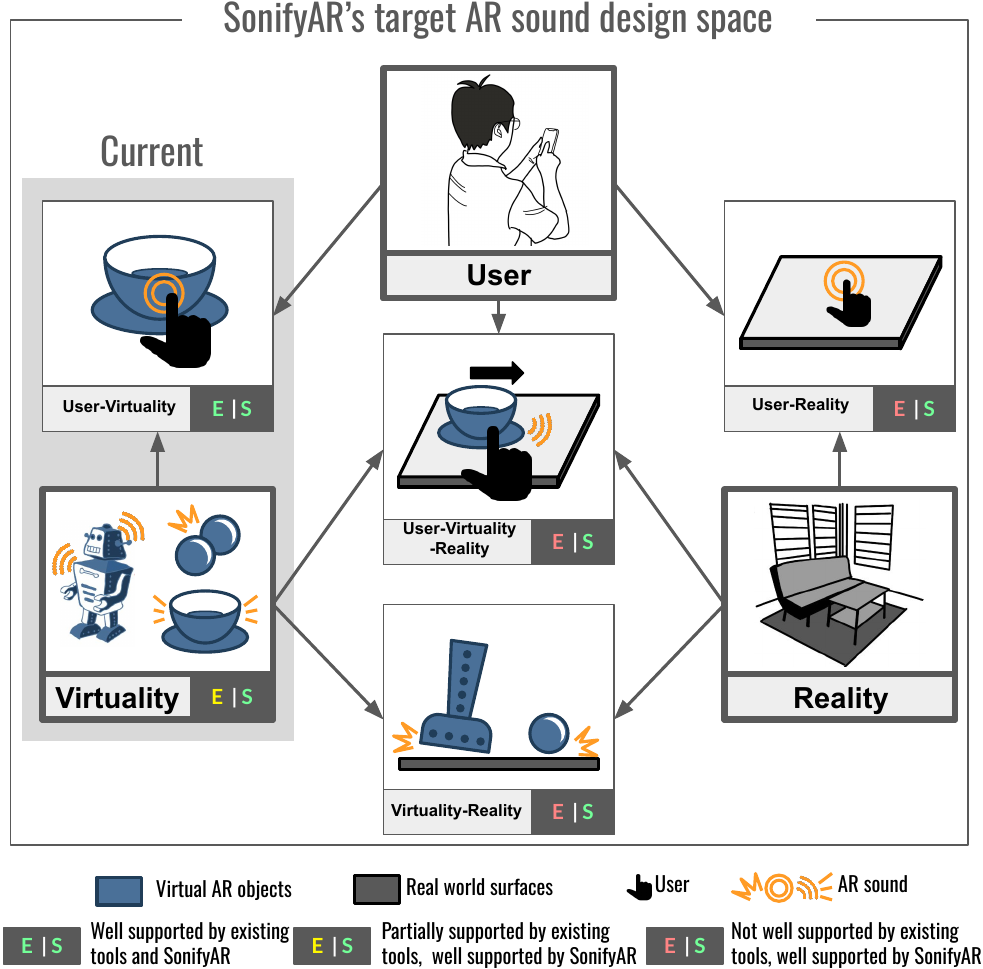}
    \caption{The sound-producing opportunities in the triad of User, Virtuality and Reality. 
    % Dashed boxes indicates the supported sound interaction space by existing tools and the expanded interaction space envisioned by our work. AR events shown in red text are examples.
    }
    \label{fig:triad}
    \Description{Sound design space shown as a hexagon between User, Virtuality and Reality. SonifyAR can support all listed sound design cases, while the existing work can only support the User-Virtuality and Virtuality cases.}
\end{figure}

% We also investigated the four AR authoring tools mentioned above for their support of the listed items and marked the support level in parentheses.

\textbf{Virtuality} (\textit{partially supported}): \textit{Virtuality} refers to sounds that accompany AR events involving only \textit{virtual objects}. This could be bound to an object's change of status (\textit{e.g.,} a notifying sound when a virtual object shows up), from an object's animated behavior (\textit{e.g.,} the mechanical noise made by a virtual dinosaur roaring), or the interaction between multiple virtual objects (\textit{e.g.,} two virtual balls clacking when they collide). Current authoring tools support status changes and animation playing as sound-initiating triggers, but do not support sound for interactions between virtual objects.

\textbf{User-Virtuality} \textit{(well supported)}: \textit{User-Virtuality} are the corresponding sounds that react to the users' actions with virtual objects (\textit{e.g.,} a virtual dog barks when users tap on or gets close to it). In all of our investigated AR creation tools, these user-initiated triggers like \textit{``Tap''} and  \textit{``Proximity Enter''} are well supported.

%\textbf{User-Virtuality} \textit{(well supported)}: When a user interacts with a virtual object, for instance tapping on a virtual dog, a corresponding sound, such as a bark, could be generated as feedback. In most AR creation tools, the trigger and its corresponding sound feedback can be predefined by specification in the authoring phase. 

\textbf{User-Reality} \textit{(not well supported)}: \textit{User-Reality} refers to sounds that accompany user interactions with the physical environment via their AR devices. This sound feedback can improve users' understanding of the surrounding environment (\textit{e.g.,} an appropriate tap sound when users tap on a real-world table surface via their phone screen). None of the AR authoring tools we investigated support user-reality actions as triggers.

%\textbf{User-Reality} \textit{(not well supported)}: This type of event refers to users interaction with the physical environment via their AR devices, even when no AR objects are present. A specific example would be a user tapping on a real-world surface through their AR device (\textit{e.g.} tapping on a wooden surface in phone screen). Sound feedback can enhance these interactions. From an accessibility perspective, such auditory cues can assist individuals with low vision, aiding them in understanding the material properties of their surroundings through AR. This sound feedback can also benefit overall immersiveness of the AR experiences, and potentially improve task performance. 

\textbf{Virtuality-Reality} \textit{(not well supported)}: \textit{Virtuality-Reality} denotes sound feedback that plays a crucial role in enhancing realism when virtual objects interact with the real-world environment (\textit{e.g.,} the crisp stomping sound of a virtual robot when it walks on a real-world glass surface). Although physics simulations between virtual models and real-world surfaces have been supported, their sound feedback remains unexplored in AR authoring tools. 

%\textbf{Virtuality-Reality:} \textit{(not well supported)}: A critical gap in AR sound authoring tools is supporting interaction between virtual and real objects. For example, when a virtual metallic robot walks on an actual glass table,  a stomping sound, resulting from the collision between metal and glass, should be produced.

\textbf{User-Virtuality-Reality} \textit{(not well supported)}: 
% When a user-initiated action involves both a virtual object and the real world environment, sound can validate this action and improve immersion. 
\textit{User-Virtuality-Reality} are the sounds that accompany user actions involving both virtual and real elements. For example, the material-aware sliding sounds when users apply a virtual scraper on different real-world surfaces (\textit{e.g.,} a wooden table, painted wall, or glass window). None of our investigated authoring tools can support the specification of this complex interaction.

\textbf{Others}: We exclude the domains of \textbf{Reality} and \textbf{User} from the enumeration since our interest lies in events where the AR system (\textit{e.g.} smartphone or AR headset) generate the sound. Thus, sounds naturally generated by user and reality (\textit{e.g.,} a user physically knocking on a wooden table) are excluded from our discussion.

%\textbf{User-Virtuality-Reality:} \textit{(not well supported)}: When a user-initiated action involves both a virtual object and the real world environment, sound can validate this action and improve immersion. For example, a user can drag and slide a virtual cup on a table surface, which can be complemented with a material-aware sliding sound (\textit{e.g., different sounds for a wood, metal, or glass table}.

Based on the above analysis and as illustrated in \figref{triad}, existing AR authoring tools fail to offer functional AR sound authoring support for many of the identified dimensions, especially those involving \textit{Reality}. SonifyAR aims to address this gap.

\subsection{Acquisition of AR Sound}
%While the above design space helps highlight the rich dimensions of AR sound, we also discover challenges for sound acquisition rooted from this diversity of sound types.
While the above design space helps highlight the dimensions of AR sound and current authoring support, below we reflect on key challenges related to the limitations of sound acquisition methods in terms of handling these different dimensions. For example, a mechanical noise of a virtual robot (virtuality) can be easily sourced from local or online sound databases. However, more specific sounds, like a virtual steel ball hitting a physical concrete wall surface or a virtual racecar jumping into a backyard pool (virtuality-reality) requires physical world understanding and material awareness. The space of possible sound effects here is near infinite. Thus, in such cases, generating sounds using text-to-sound models is preferable. Conversely, certain highly recognizable sounds, like a dog barking, pose difficulties for current generative models, often resulting in outputs that are unsatisfactory and noisy.
% Some sounds, like dog barking, are highly relatable thus the generated sound, which generally is noisy and subpar in quality, cannot fulfill user expectation. Whereas some highly imaginary sounds like dinosaur roar, can be fulfilled by generated assets. 
The complexity in user needs and trade-offs in different methods suggests that the combination of both generation and retrieval could serve as a practical solution. %<Any citation or example that have similar solution?>

%To facilitate sound retrieval functionality, similar to what is offered by the existing AR authoring tools like Aero and Reality Composer, we select a collection of popular sound effect snippets as our local datasets. Additionally, we ultilize FreeSound \cite{freesound} as our online retrieval database. Both the local and online databases can be expanded to meet customized user needs. For generative sound capabilities, we choose AudioLDM \cite{liu2023audioldm} for its state-of-the-art performance, as well as its capability in both generating and transferring sound. This enables us to create sound from scratch and modify existing sound snippets for user editing. Together, these approaches form a comprehensive suite that expands our capacity to accommodate a broader spectrum of sound interactions than any single method could achieve.

\subsection{Event Representation}
To combine generation and retrieval methods, one challenge is how to condition this suite of different sound acquisition methods.
%Another challenge is conditioning the suite of sound acquisition methods to produce accurate sound outputs. 
While visual formats such as images \cite{10096023} and videos \cite{Zhou_2018_CVPR,ghose2022foleygan} have been widely used as inputs for sound generation models, they fall short in fully capturing AR context information, particularly user actions. To overcome this, we explore the use of text as a universal representation for sound acquisition inputs, which has the following benefits. First, text can sufficiently and precisely convey necessary context information and the specifics of sound-generating events (\textit{e.g.,``User slides a ceramic cup on a wooden surface''}). Second, given that the AR experience operates within a hardware-software system with multi-modal sensors (\textit{e.g.,} camera, IMU, GPS), we can leverage this system to monitor and log events within the AR experience and summarize them into descriptive text. Third, recent advancements, as demonstrated by works like \textit{HuggingGPT}~\cite{shen2023hugginggpt} which uses a LLM to control AI tasks, showcase the practicality of using LLMs as controllers to process text for sound generation.

\section{SonifyAR Implementation}
\label{sec:implementation}
\begin{figure*}[t]
    \centering
    \includegraphics[width=\textwidth]{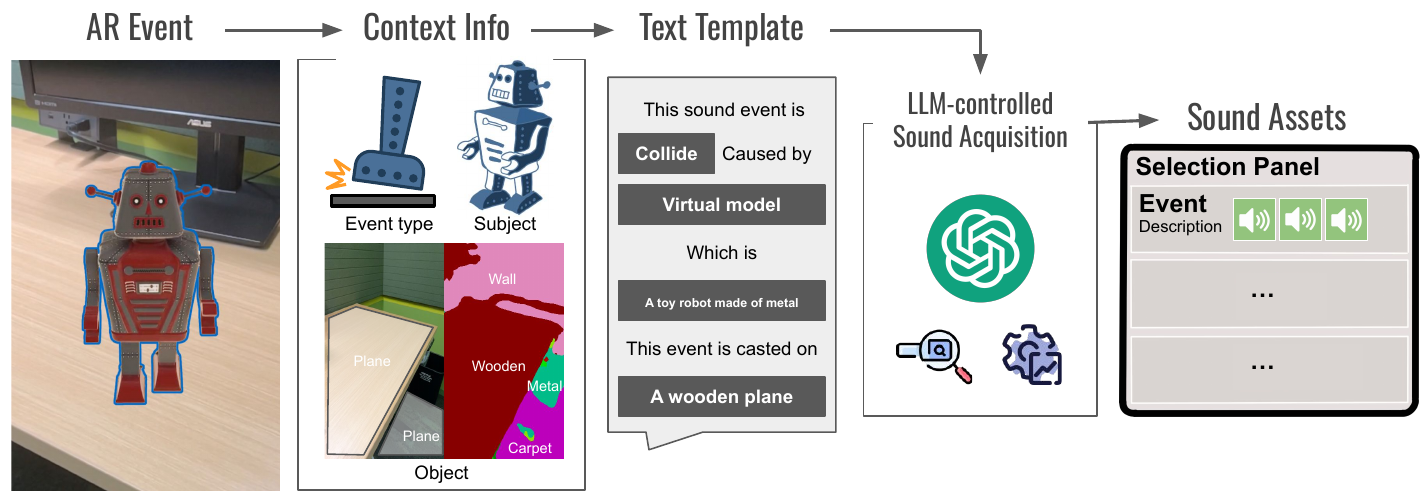}
    \caption{Overview of the pipeline of SonifyAR. Our system monitors and logs context information of AR events, which includes the event type, the subjects and objects (virtual or real-world), and the attributes of the involved elements like their materials. This information is compiled into a text template and then processed by our LLM controller to acquire sound assets. The results are subsequently presented in our selection panel. }
    \label{fig:pipeline}
    \Description{A pipeline figure showing SonifyAR's technical process. From left to right: a virtual toy robot on a wooden desk; several smaller figures showing the context information; a text box containing the context; three icons showing the LLM-controlled sound acquisition process, including openAI logo, a search icon, and a gear icon; a simplified UI panel for sound assets selection.}
\end{figure*}
Building on the above design space, we introduce SonifyAR, a novel PbD sound authoring framework that uses context recognition and generative AI to create personalized, context-sensitive sounds for AR interactions (\autoref{fig:pipeline}). SonifyAR consists of three primary components : (1) \textit{Event Textualization}: Every user action is recognized as an AR event, and the context of these events is transformed into textual descriptions. (2) \textit{Sound Acquisition}: An LLM-controlled sound acquisition process that utilize four methods to produces sound effects based on the context. (3) \textit{User Interface}: An interface that allows users to experience the AR scene and provides the capability to view, modify, and test sound effects for AR events.

\subsection{Event Textualization}
\label{subsection: event textulization}
% \begin{figure}[h]
%     \centering
%     \includegraphics[width=0.9\textwidth]{Figures/event_temp.png}
%     \caption{Event recognition.}
%     \label{fig:event}
% \end{figure}

In SonifyAR, we utilize a PbD authoring framework to capture potential sound-producing AR events and compile the context information into text. When users interact with the AR space, SonifyAR captures AR events that can lead to sound feedback. Here we define an AR event as a user action and its subsequent result (\textit{e.g.,} when a user taps a virtual object to trigger its animation). For each AR event, we describe it with an event type (\textit{e.g.,} tapping an object ), action source (\textit{e.g.,} user or virtual object), and action target (\textit{e.g.,} virtual object or real-world plane). Information about the involved parties, such as virtual objects or real-world planes, are also included in the context information. This collected event data is then transformed into text with a template.

\begin{figure}
    \centering
    \includegraphics[width=0.5\linewidth]{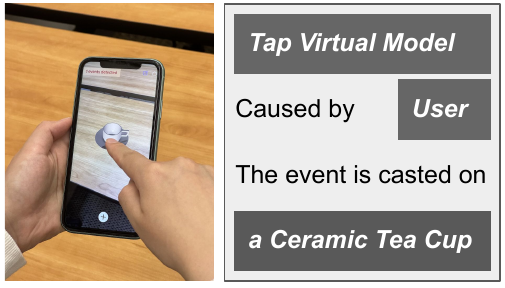}
    \caption{SonifyAR's event textualization. Left: a user tapping on a virtual cup through the SonifyAR interface; Right: the textual context information extracted by SonifyAR's internal PbD framework.}
    \label{fig:enter-label}
    \Description{Left: a human user holding a phone, while tapping on the screen with another hand. There is a virtual ceramic tea cup on the screen; Right: a text box containing the context information.}
\end{figure}

\paragraph{Event Types.}
To support a broad range of AR sound interactions, we implemented six event types: 

% It is essential to note that these six events serve primarily as proof-of-concept. Users can easily adapt our methodology to accommodate new event types not explicitly defined in our framework.

\begin{enumerate}
    \item \textit{Tap Real World Structure:} A user taps on a real-world structure through the AR interface.
    \item \textit{Slide: }A user holds the virtual object and slides it on a real-world surface.
    \item \textit{Collide:} A virtual object collides with another virtual object or with a real-world surface. We implemented a specialized collision detection mechanism as introduced in \autoref{subsection:Implementation}
    \item \textit{Show Up: }A virtual object shows up in the AR experience.
    \item \textit{Tap Virtual Objects:} A user taps on a virtual object through the AR interface.
    \item \textit{Play Animation:} A virtual object plays an animation.
\end{enumerate}

\paragraph{Scene Context Understanding.}
SonifyAR utilizes ARKit \cite{apple-arkit-doc}'s plane detection functionality \cite{apple-arkit-plane-tracking} to assess the surrounding environment. Since most AR experiences are anchored to planes, the detected plane information serves as a crucial context. To enrich such information, we employ the \textit{Deep Material Segmentation} model \cite{upchurch_dense_2022} to segment the scene and identify the material of planes (\autoref{fig:pipeline}), which helps produce realistic sound effects when an AR event involves this plane. Currently, we support six materials: wood, carpet, concrete, paper, metal, and glass. Surfaces that do not fall into one of these six categories are labeled as \textit{unknown surface}. %In the future, we can use a more advanced model to accommodate a broader range of materials.

\paragraph{Virtual Object Understanding}
Virtual object semantics are also crucial for sound generation. To ensure sound assets align with the virtual object's material and its state, we collect a text description for all the virtual objects (\textit{e.g.,} ``\textit{This model is a toy robot made of metal.}'') and their animations (\textit{e.g.,} ``\textit{A toy robot walks.}''). These descriptions can be provided by the asset creator or, if necessary, we prompt the user to add relevant details. Any AR event involving the virtual object or its animation will incorporate these text descriptions.

\paragraph{Event-to-text}
To aggregate the multi-source context information described above, we employ a simple text template that will feed into an LLM-based backend: ``\textit{This event is [Event Type], caused by [Source]. This event casts on [Target Object]. [Additional Information on Involved Entities].}'' \textit{Event Type} is the type of event described by the aforementioned type names; \textit{Source} refers to the subject of the event, which could be the user when the event is directly triggered by user, or virtual object when it interacts with real-world environment; \textit{Target Object} is the object of the event (\textit{e.g.,} the plane that gets tapped on or the animation that gets played.). Finally, \textit{Additional Information on Involved Entities} include details that further elucidate the source object, the triggered event, and the target object. Examples include material descriptions and animation details. %Users are asked to provide this data within the metadata of each 3D model.
We leave the corresponding entity field blank when events are missing a target object or additional information.

During user interactions, the system actively monitors and logs events happening in AR space. The context information is fetched and plugged into the text template, crafting a coherent description that reflects the user's interaction within the AR environment.

% \paragraph{Pre-defined Event}
% As the AR scene begins to load, our system starts scanning the memory to identify which AR models have been loaded and what types of events have been designated for each AR object (e.g., appear/disappear, animation). Each event is then described as a natural language sentence using a template. \cmt{design and explain the template}. Once the template is populated with specific system components, the constructed sentence is sent to a language model to determine if the event requires sound. If a sound is necessary, we register this event in the "sound-producing event" list, and it is then queued for processing by the sound production component. Please refer to \cmt{Fig X} for an illustration.

% \paragraph{User Interaction Event}

\subsection{Sound Acquisition}

\begin{figure}[h]
    
  \centering
  \includegraphics[width=\linewidth]{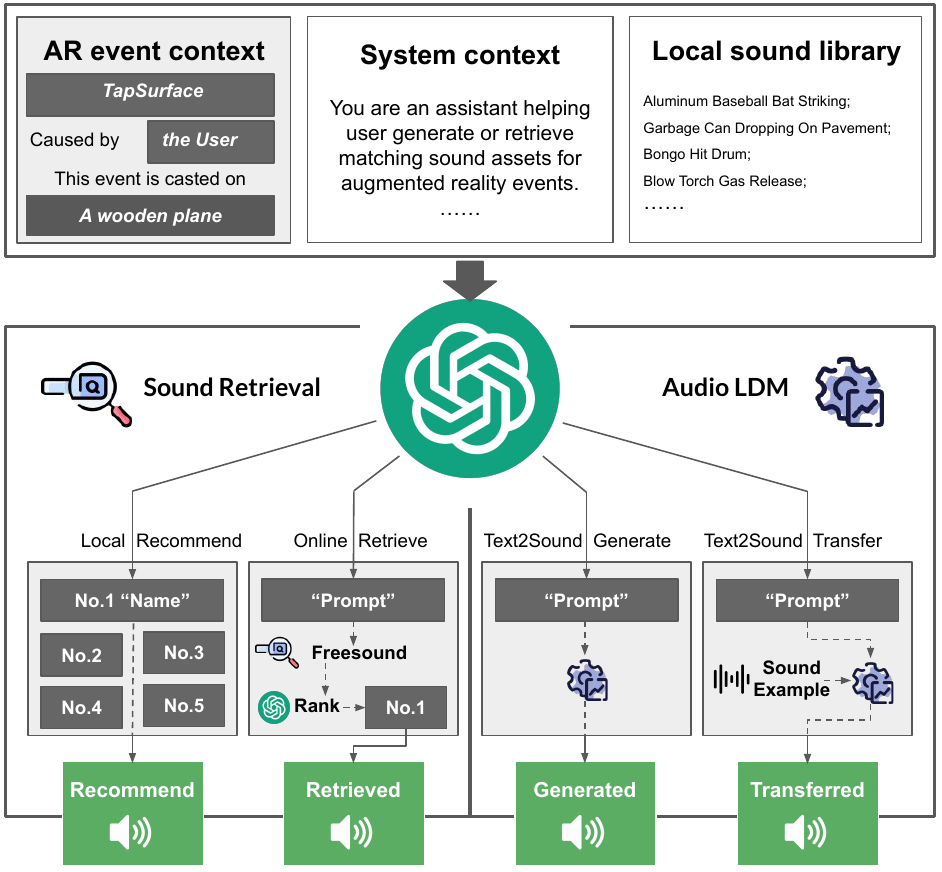}
  \caption{SonifyAR's sound acquisition pipeline.}
\label{fig:generation}
\Description{A detailed diagram showing SonifyAR's sound acquisition process. Top shows the context information fed to the LLM; Middle shows a big openAI logo, indicating the LLM; Bottom shows four sound acquisition pipeline and their respective outputs.}
\end{figure}
%Need a figure here to explain the authoring pipieline, including the text content sent to and from the LLM.

%In a traditional AR authoring system, users access a provided sound library to choose appropriate sound assets. If no suitable sound can be found in the library, users may search for sounds online or record sound by themselves. Once an ideal sound effect is identified, it can be added to the sound library and associated with a specific event. This sound authoring process can be labor-intensive and challenging, posing barrier to novice AR creators. With SonifyAR, we aim to simplify and automate this process using LLM and generative models to reduce the time and learning curve in AR sound authoring. 

SonifyAR utilizes GPT4\cite{openai2024gpt4} to automatically retrieve or generate context-matching sound assets of an AR event. Inspired by \textit{HuggingGPT}~\cite{shen2023hugginggpt} and \textit{Visual ChatGPT}~\cite{wu2023visual}, we utilize the LLM as a controller of multiple sound authoring methods. The LLM takes the text description of the event as input and replies with commands for multiple sound acquisition methods (\autoref{fig:generation}). At our current stage, we support four major sound authoring methods: local recommendation, online retrieval, \textit{text2sound} generation, and text-guided sound transfer. All sound authoring processes listed below operate concurrently in the backend. This ensures an uninterrupted AR experience.

\textbf{Local Recommendation.} The LLM can recommend sound assets stored in the local database based on semantics in the event description. Similar to other AR authoring tools, we collect a set of sound effects from \textit{Adobe Audition}'s library\cite{adobe-audition}, each labeled with a descriptive filename, like ``\textit{Crash Aluminum Tray Bang}'' or ``\textit{Liquid Mud Suction}''. The entire catalog of sound filenames is provided to LLM. When given the event context, the LLM recommends the top five most suitable local sound effects based on their filenames. It then returns the selected sound effect in the format of \texttt{method1recommend:FILENAME}, where  \texttt{FILENAME} is
replaced by the actual filename. Upon receiving, SonifyAR parses the filename and adds the top corresponding sound as one of the sound options for the event. Users can also long-hold options in the UI to reveal other top recommendation results.

\textbf{Online Retrieval.} We also expand our retrieval capability with an online sound asset database called \textit{FreeSound} \cite{freesound}. We use the FreeSound API, which returns an N-best list of matching sound effects based on a given query.
The queries are condensed versions of full event descriptions generated by the LLM, returned in the format \texttt{method2retrieval:PROMPT}, with \texttt{PROMPT} replaced by the specific search query. The returned results are JSON strings. We select the top five matches from the entire set of sound effects returned by the API, which are then downloaded and presented to the user as sound options. 

\textbf{Sound Generation.} Beyond recommendation and retrieval methods, we also use the text prompt to generate custom sound effects using an audio diffusion model \textit{AudioLDM}~\cite{liu2023audioldm}. We ask the LLM to compress the event text description into a shortened generation prompt: using the format \texttt{method3generation:PROMPT}, where the \texttt{PROMPT} would be replaced with the generation prompt. Upon receiving such a command from the LLM, SonifyAR sends the prompt to the AudioLDM model, requesting \textit{text2sound} generation. These newly generated sounds are then shown in the UI as sound options. 

\textbf{Sound-style Transfer.} Besides text2audio generation, \textit{AudioLDM} is also capable of performing text-based sound style transfer. Specifically, for events like tapping, sliding, or colliding, instead of generating a new sound from scratch, SonifyAR uses a default sound effect and initiates a style transfer operation with a text prompt provided by the LLM (\textit{e.g.,} transfer a general tap sound to tapping on glass). This approach allows the output sounds to match the length and rhythm of the input sounds, so that they can be well-timed with actions. Furthermore, when users wish to further modify any provided sound assets, SonifyAR offers text-based style transfer as a fine-tuning and customizing option.

% We use three types of technique to generate sound given a text description.
% 1. retrieve on collection (LLM recommendation based on title)
% 2. retrieve on online resources (we use freesound)
% 3. text-to-sound generation (AudioLDM), explain details in implementation (if else).

\subsection{User Interface}
\label{subsec:UI}
%Need an interface figure
\begin{figure*}[h]
    
  \centering
  \includegraphics[width=\linewidth]{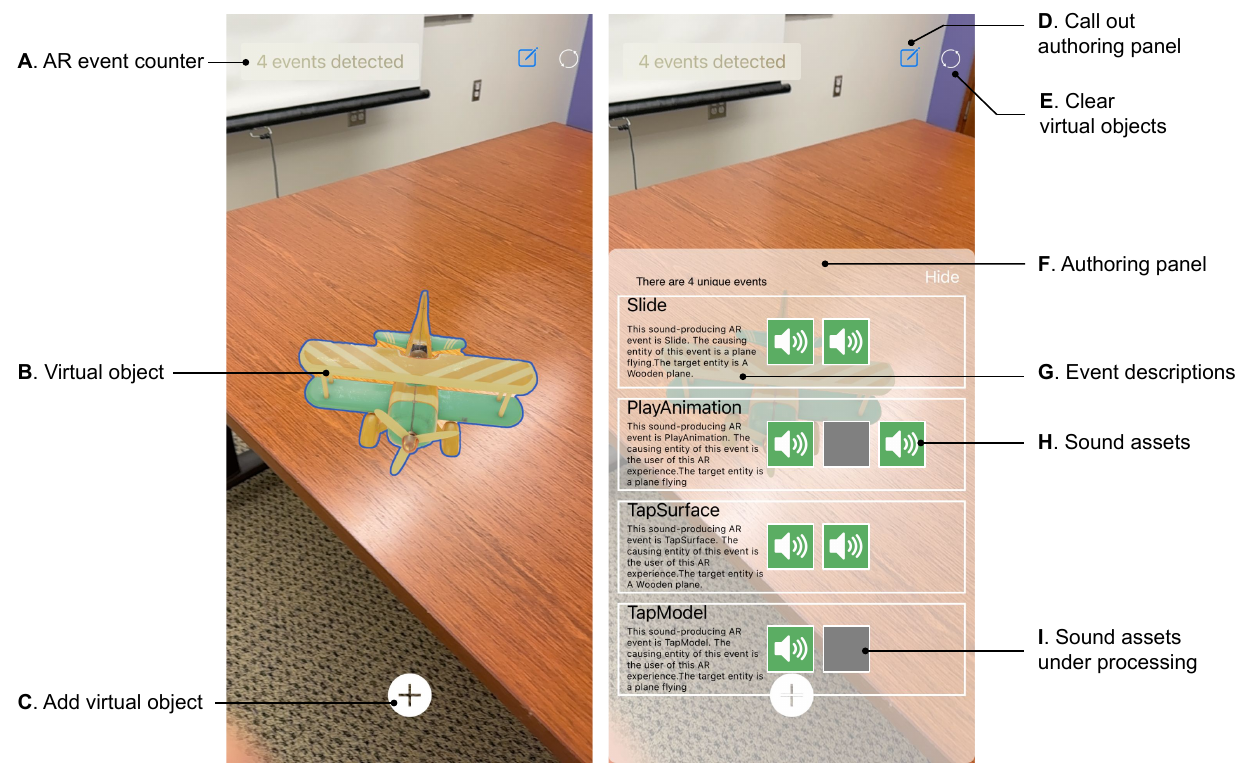}
  \caption{SonifyAR's authoring interface. Left: SonifyAR's phone-based AR interface. Right: SonifyAR's authoring panel.}
\label{fig:interface}
\Description{Two UI screensots. Left shows a virtual toy plane placed on a real-world wooden desk; Right shows the same interface with a half-transparent call-out panel on the lower half of screen, containing the acquired sound assets for AR events.}
\end{figure*}
%Our system adopts a PbD authoring framework that enables users to author sound effects interactively while engaging with the AR environment. This approach offers a more intuitive authoring experience by helping users understand the relationship between their interaction behaviors and the potential generated sound effects.

As a PbD authoring framework, the SonifyAR application invites users to explore freely with an interactive AR experience (\autoref{fig:interface} left). Users can interact with the scene, performing actions like moving virtual objects or interacting with the real-world surfaces. When an AR event is detected, a text label appears on the top-left screen to inform the user (\autoref{fig:interface}A) and the sound acquisition component is activated to generate candidate sound effects for the detected events. By clicking the ``\textit{authoring panel}'' button (\autoref{fig:interface}D), users can prompt an editing interface (\autoref{fig:interface}F) that overlays the AR scene. The UI includes all detected AR events (\autoref{fig:interface}G) and the corresponding generated sound assets (\autoref{fig:interface}H). Users can click on a sound effect to preview and double click to select and activate it. After confirming their choices, users can hide the authoring panel to resume the AR experience and test the selected sounds. The editing interface can always be re-activated to modify choices.

Additionally, users can long press a sound asset to call out a menu with a suite of exploratory options. For recommended or retrieved sounds, the menu includes an option to list all other sound assets in the top five recommendations. For all sound assets, the menu provides a ``\textit{style transfer}'' and a ``\textit{generate similar sounds}'' feature, enabling users to style transfer audio effects or to generate similar sounds based on the selected sound. When this feature is selected, users can type a simple text prompt to guide the sound generation process. These choices can be iterated upon or used to explore new sound variations.

\subsection{Technical Implementation}
\label{subsection:Implementation}
SonifyAR is built with Apple's ARKit in Swift 5.7 and Xcode 14, and runs as an iOS application on iPhones. Due to the enhanced performance of plane detection on devices with LiDAR (available in iPhone's Pro lineup starting from iPhone Pro 12), we developed and tested the app on an iPhone 13 Pro Max running iOS 16. We implemented the collision detection between virtual objects and planes with ARKit's physics simulation \cite{apple-scenekit-physics-simulation}. To enable collision detection for animated virtual object parts (\textit{e.g.} the stomping of a robot foot on the plane), we bind colliders to the joints of the virtual object's animated bones.

For the LLM-based backend, we use GPT-4.0 \cite{openai2024gpt4}. The text prompt is shown in \autoref{appendix}. For material segmentation, we use the \textit{Dense Material Segmentation} (DMS) model 
\cite{upchurch_dense_2022} for its detailed material labeling, fast processing time, and high accuracy. For sound generation, we use the \textit{AudioLDM} model \cite{liu2023audioldm} due to its state-of-the-art performance and versatility in text2sound generation and text-guided sound2sound transfer. All models are hosted on a server with an NVIDIA RTX 4080 GPU and CUDA 12.2.

% Implementation details on what exact model/algorithm used for each component, and our device used for testing.

% We want to emphasize that our system is focusing on sound authoring, so we need to take authored AR scene as input because creating AR scene is beyond the scope of our system. This makes our system to be use as an add-on to any other AR authoring tools like Aero. Once the AR scene is authored and launched, we initiate our system to generate sound effect to the scene.

% Some overview of the pipeline, with the diagram.
% \subsection{Context as text}
% Collect context information from multiple sources and models, turn them into text.
% \subsection{Sound from context}
% With the text containing context information, we could generate/retrieve/transfer sound to adapt to context.

% We use sound generation model, sound retrieval API, and recommendation of LLM.
% \subsection{User interaction}
% We consider user intention as part of the context information. Thus engaging user in a interaction pipeline is necessary. user could express their design intention and preferences over multiple choices in an iterative interaction based on language and UI taps.

\section{User Study}
To examine the usability and authoring performance of the SonifyAR system, we conducted a usability study across eight participants, who used the SonifyAR application in a fixed experimental setting and provided ratings for the system.

\subsection{Participants}
We recruited eight participants (six male and two female, aged between 26 and 33)\rev{ via snowball sampling at Adobe with varying experience in AR. }Out of the eight participants, six had prior AR experience, and three had specific experience in AR authoring tools.

\subsection{Procedure}
The usability study was conducted in a small meeting room and had four parts: (1) We first introduced our study goal, collected demographic and background information. (2) We then demonstrated SonifyAR with an example AR experience. During this phase, participants had the opportunity to ask any questions regarding SonifyAR's functionalities. (3) Afterwards, participants were asked to independently explore the SonifyAR app. We provided a walking robot model which could be added to indoor surfaces. This model starts walking when tapped. For consistency, we asked all participants to explore this same AR asset and use our automatic sound authoring pipeline to create sound effects with the model until they were satisfied with the results. The entire usage process was screen-recorded and the user operations were logged. (4) Finally, we sought participant feedback regarding the usability, helpfulness and technical performance of SonifyAR. Participants provided reasoning and insights to support their answers. The questions and rating results are shown in \autoref{fig:stats}.

\subsection{Results}
All participants were able to complete the AR sound authoring task using SonifyAR without difficulty. On average, the authoring process took 406 seconds (\textit{SD=}137s),\rev{ and the entire study took 35 minutes}. Participants tested an average of 56 (\textit{SD=}10.5) sound assets, with an average of 6.4 sounds (\textit{SD=}1.7) assigned to the authoring results.

\begin{figure}[h]
    
  \centering
  \includegraphics[width=\linewidth]{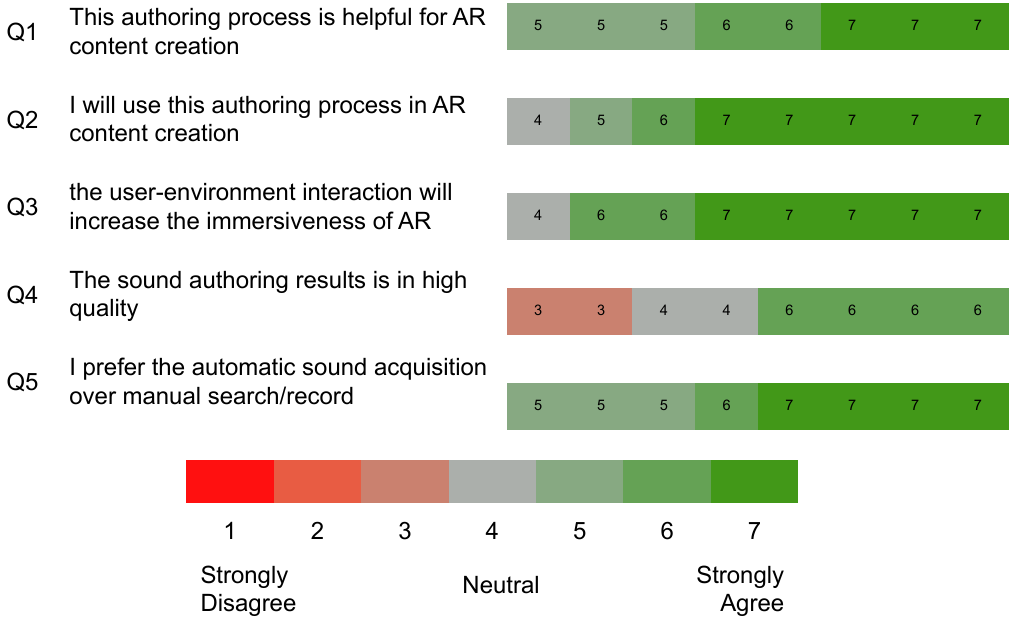}
  \caption{Likert questions and results of our usability study}
\label{fig:stats}
\Description{A horizontal chart showing the likert question results with color and numbers.}
\end{figure}

Participant feedback was also largely positive. All eight participants expressed favorable impressions of the tool. They highly agreed that the SonifyAR tool would be helpful to AR authoring process (Q1, \textit{avg=}6 out of 7, \textit{SD=}0.93). There was also a general willingness for using SonifyAR in their own AR creation practice (Q2, \textit{avg=}6.3, \textit{SD=}1.16). Participants agreed that the sound interaction involving real-world surfaces will improve the immersiveness of AR experiences (Q3, \textit{avg=}6.4, \textit{SD=}1.1). Additionally, they preferred the automated sound authoring process over their prior manual search experiences for sound assets (Q5, \textit{avg=}6.1, \textit{SD=}1.0). The only area for improvement was the quality of the generated sound (Q4, \textit{avg=}4.75, \textit{SD=}1.39). However, this could be enhanced with the introduction of a more advanced sound generative model, which can be easily integrated into our system. The full list of questions and the respective Likert results can be viewed in \autoref{fig:stats}.

\rev{Besides the ratings, participants also provided suggestions, especially on the UI and usability. For example, \textit{``I really want more usability features that can help communicate these events''}, and \textit{``I hope there are more associations between the AR events and the sound generation happening in the backend, like some real-time audio hints''}. }

\section{Applications}
To further explore and demonstrate the potential of our approach, we authored five AR scenarios with SonifyAR. See the supplementary video for additional details and an audio-visual demonstration of the resulting SonifyAR sound effects.

\subsection{Education}
\begin{figure}
    \centering
    \includegraphics[width=0.95\linewidth]{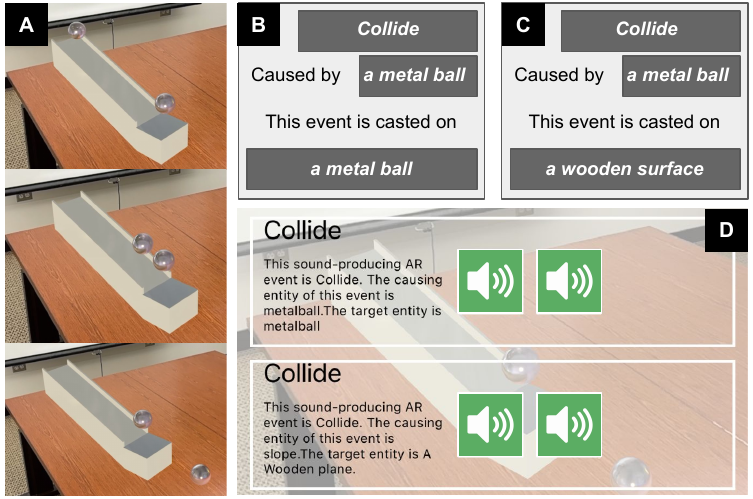}
    \caption{An AR physics experiment sonified by SonifyAR. (A) The slope and metal ball model and simulation process. (B), (C) Context information of two collisions in the simulation. (D) The acquired sound assets for the collisions.}
    \label{fig:applicationA}
    \Description{A figure showing an AR physics experiment and SonifyAR's authoring results. A: a virtual slope with two metal balls on it. The top ball slide down and collide with the lower one. Then the lower one drop down to the table while the top one remain on slope. B and C: the context information of the two collide events; D: the resulted sound assets shown in UI.}
\end{figure}

%Use bumping balls to show how can it improve physics education.
AR has been widely explored in STEM education (\textit{e.g.,}~\cite{sirakaya2022augmented, Kang_ARMath_CHI2020, Kang_PrototypAR_IDC2019, Kang_SharedPhys_IDC2016, Suzuki_RealitySketch_UIST2020, Chulpongsatorn_AugmentedMath_UIST2023}). As existing educational applications try to simulate and visualize physical or chemical phenomenon in AR \cite{radu2019can,irwansyah2018augmented}, SonifyAR can improve the blending between the real environment and virtual content through immersive, appropriate sound effects. To showcase such a possibility, we implemented an AR physics experiment using SonifyAR. To help illustrate one of Newton's laws of motion--the conservation of momentum---we created an AR scene with a downward ramp and two metal balls: one at the top and the other at the bottom (\autoref{fig:applicationA}). Upon initiating the experiment, the top ball rolls down the slope and collides with the bottom ball. This collision causes the bottom ball to move forward and eventually fall, while the top ball comes to a stop. With SonifyAR, collision events are automatically detected and material-specific sound effects, such as the clashing of two metal balls or a metal ball dropping onto table, are seamlessly generated.

%SonifyAR can help educational AR applications better simulate physical phenomenons. Existing AR education apps all focus on visual virtual content and cannot show the physical interaction between real world and virtual objects. SonifyAR can help generate material-aware sound that makes virtual educational content more immersive and realistic.
\subsection{Accessibility}
%Use Little robot to show how can it help low vision users locate objects
AAR (Audio Augmented Reality) has been explored as accessibility assistance for blind or low vision people~\cite{ribeiro2012auditory,coughlan2017ar4vi}. By sonifying real world environments and virtual contents, blind or low vision (BLV) people can better interpret visual information in both reality and virtuality. We envision SonifyAR assisting AR accessibility in three ways. First, by supporting AR sound authoring, SonifyAR encourages creators to add sound effects in AR experiences, which can help BLV users interpret visual content. Second, by providing context-aware AR sound in 3D audio, people with low-vision can better navigate virtual objects in the AR environment~\cite{ribeiro2012auditory}. Third, by enabling user to interact with real-world surfaces via AR (\textit{e.g.} tapping on a real world surface), user can explore the surrounding space via the AR interface.

\begin{figure}
    \centering
    \includegraphics[width=0.6\linewidth]{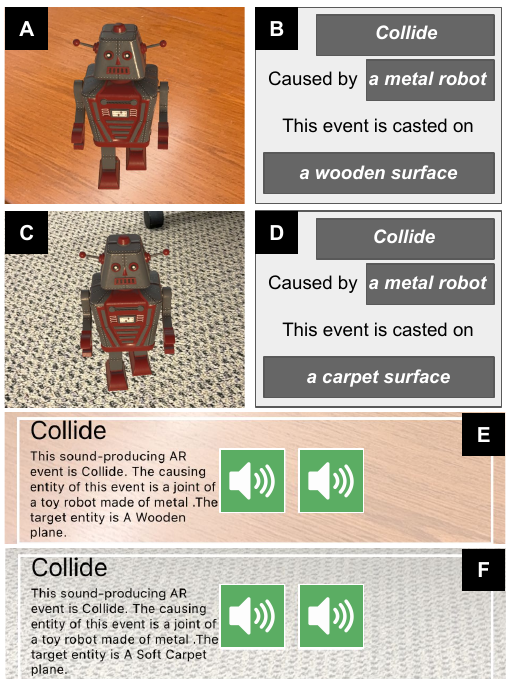}
    \caption{Robot walking on different surface with different sound. (A), (B) Robot walking on a wooden surface, and the corresponding context information. (C), (D) Robot walking on carpet surface, and the corresponding context information. (E), (F) The acquired sound assets.}
    \label{fig:applicationB}
    \Description{A figure showing an AR robot walking and SonifyAR's authoring results. A: a virtual robot walk on wooden table. B: the context information of the walking; C: a virtual robot walk on carpet surface. D: the context information of the walking; E and F: the resulted sound assets shown in UI.}
    
\end{figure}

To showcase potential accessibility benefits, we use a toy robot 3D model (\autoref{fig:applicationB}). As the robot virtually walks from one physical surface to another (\textit{e.g.,} from a wood floor to a carpeted floor), the auto-generated sound effects change appropriately. In this way, a BLV user can better perceive the location and activity of the robot as it is walking. 

\subsection{Sonify Existing Apps}
%Use the AR camera app to show the results
%Lots of existing AR application lack sound feedback (any citation? Or should we quickly test out some apps to support this point). If the SonifyAR pipeline can be adopted at SDK level, it would be possible to provide sonification using the automatic sound acquisition results.

Due to the extensibility of the text template used in SonifyAR's sound acquisition pipeline, it can be easily implemented into existing AR apps. For example, if implemented at the SDK level (\textit{e.g.} ARKit and ARCore), SonifyAR can automatically capture textual description of AR events and provide sonification using the automatic sound acquisition results.

\begin{figure}
    \centering
    \includegraphics[width=0.7\linewidth]{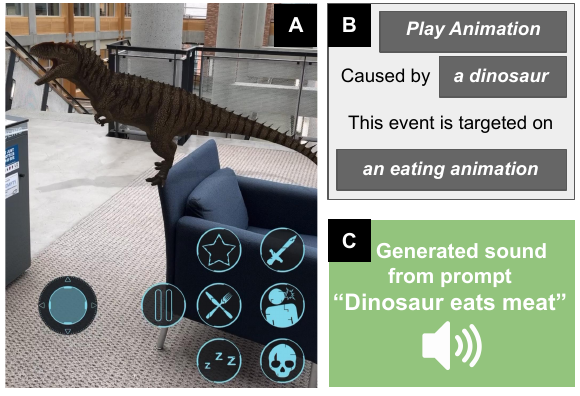}
    \caption{AR dinosaur sonified by SonifyAR. (A) AR Camera app interface with a virtual dinosaur playing eating animation in the scene. (B) The context information of the scene. (C) Text prompt and sound output from SonifyAR's pipeline.}
    \label{fig:applicationC}
    \Description{A figure showing an dinosaur roaring and SonifyAR's authoring results. A: a virtual dinosaur roar in an AR scene; B: the context information of the roaring; C: the resulted sound asset.}
\end{figure}

We explore this application possibility with a Wizard of Oz (WoZ) prototype on a phone-based AR application called \textit{ARVid} \cite{arvid}, which is a downloadable AR app that puts 3D animated objects into real world environment. We selected a dinosaur animation as an example (\autoref{fig:applicationC}) and use WoZ to simulate the direct integration of the SonifyAR pipeline into ARVid app. By manually feeding the context information (\textit{e.g. ``The dinosaur is eating''}) into the SonifyAR pipeline, we generated sound assets that well-match the 3D animations. 

\subsection{Using SonifyAR on an MR Headset}
%One recent trend in AR, or termed by Meta Quest and Apple Vision Pro, mixed reality (MR) is closely engaging virtual content around real world environment. SonifyAR can help strengthen the object permenance of the virtual windows bu adding sound feedback when these virtual windows intersect real-world surafces.

As recent AR/MR headsets become increasingly popular, manufacturers have established safety guidelines, such as setting up safety zone and clear up indoor spaces. We envision SonifyAR contributing to this topic by hinting users of the existence of real-world entities with material-aware sound effects when application windows or virtual object intersect with real world surfaces or objects. %This sound 

\begin{figure}
    \centering
    \includegraphics[width=0.8\linewidth]{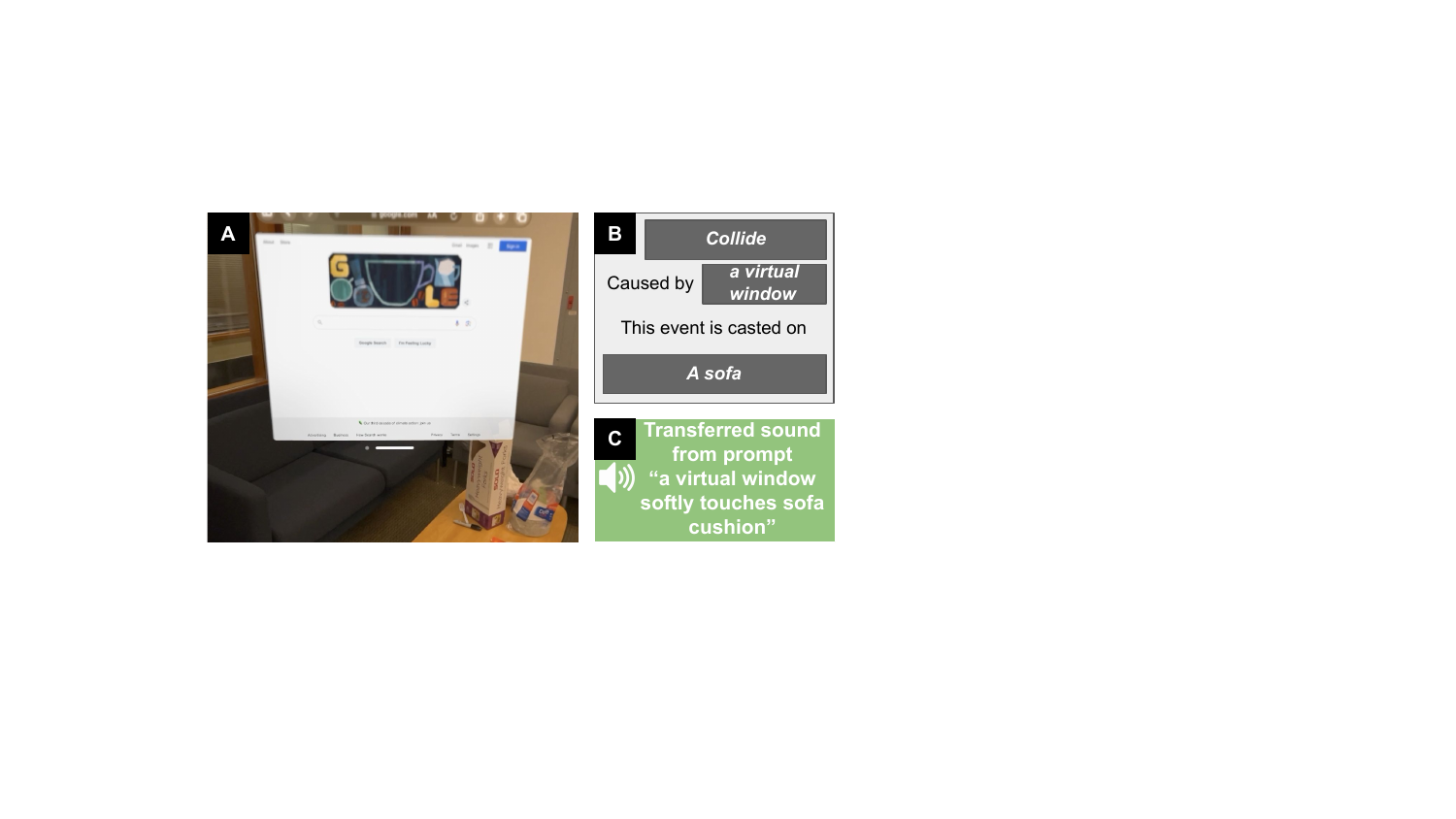}
    \caption{Sonification of Vision Pro's window operation. (A) User place a virtual window on sofa. (B) Context information of this action. (C) Text prompt and sound output from SonifyAR's pipeline.}
    \label{fig:applicationD}
    \Description{A figure showing Vision Pro's interface and SonifyAR's authoring results. A: a virtual browser window rests on a sofa in Vision Pro's interface; B: the context information of the collision; C: the resulted sound asset.}
\end{figure}

We implement a WoZ prototype based on Apple Vision Pro's user interaction (\autoref{fig:applicationD}). When using the Vision Pro, a user can actively adjust the position of application windows by hand gestures. SonifyAR can enhance this interaction by generating AR sounds when virtual windows are placed on or intersect with indoor surfaces like walls, desks, and floors. By processing these interactions and applying sound transfers with text prompts such as \textit{``a virtual window bumps into a wall''} and \textit{``a virtual window collides with a carpet''}, SonifyAR aims to improve the permanence of virtual windows and increase awareness of surrounding objects as users navigate through virtual windows.

\subsection{Augmenting AR Authoring Processes}
%Use apple's reality composer to showcase possibility
Besides serving as a standalone PbD authoring experience, SonifyAR could improve existing AR authoring tools. As mainstream AR authoring tools like Adobe Aero and Apple Reality Composer already represent their AR interaction as text like \textit{``Tap \& Play Sound''} and \textit{``Proximity \& Jiggle''}, these specifications can be fed into SonifyAR's text-based sound authoring pipeline. In this case, SonifyAR works as a plugin and provides matching sound assets for AR interactions.

\begin{figure}
    \centering
    \includegraphics[width=1\linewidth]{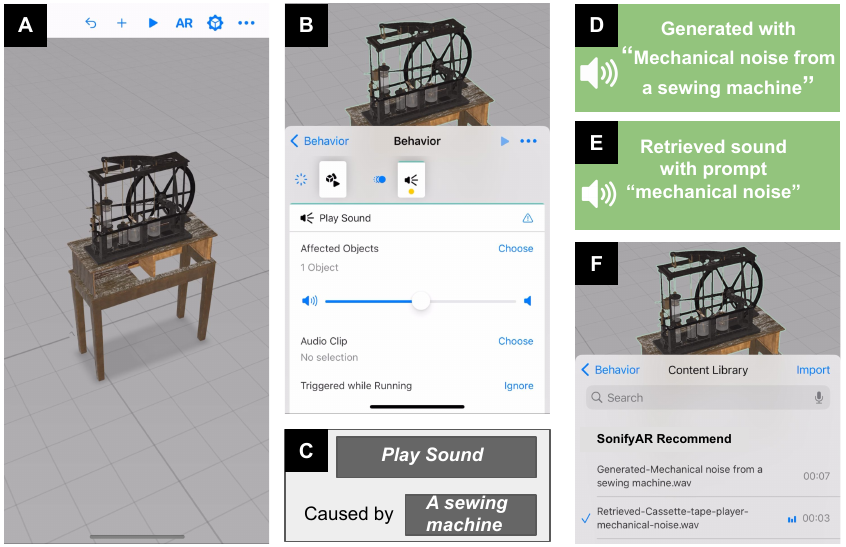}
    \caption{WoZ example of SonifyAR being applied in Reality Composer's authoring process. (A) A machine model in Reality Composer's authoring interface. (B) Reality Composer's interface for AR sound. (C) Context information of the event in B. (D), (E) Two sound assets generated and retrieved with SonifyAR's pipeline. (F) The acquired sound being used as recommended options.}
    \label{fig:applicationE}
    \Description{A figure showing Reality Composer's interface and SonifyAR's authoring results. A: Reality Composer's mobile interface with a sewing machine in the canvas; B: Reality Composer's UI for authoring the "play sound" action of the virtual object; C: the context information of the "play sound" action; D: SonifyAR's generation result as a sound asset; E: SonifyAR's retrieval result as a sound asset; F: the WoZ interface of Reality Composer showing SonifyAR's sound assets.}
\end{figure}

We implemented an additional WoZ prototype using Apple's Reality Composer and a 3D model of an animated machine (\autoref{fig:applicationE}). When authoring the \textit{``Play Sound''} behavior of this model, context information like the model description, triggering condition, and animation description, can be fed into SonifyAR's text-based authoring pipeline. The resulting retrieved/generated sound assets appear at the top of the authoring UI as \textit{``SonifyAR Recommended''}. See supplementary video for more details.

\section{Discussion and Future Work}
In this paper, we designed and implemented an AR sound authoring pipeline that automatically generates sound assets based on context information. We evaluated and identified opportunities in the AR sound interaction space, and implemented a custom LLM+Generative AI pipeline for sound acquisition. We tested the pipeline with a usability study and through a "proof-by-demonstration" across five application scenarios.

%\subsection{Sound quality}
%Based on our usability study results (\autoref{}), participants are generally satisfied with the system's usability and technical performance. However, according to multiple comments, the quality of the acquired sound can still be improved, especially for the text-to-audio generated sounds.
%We argue that although the current sound generation quality is not perfect, the text-to-audio generation is a new and popular research field with countless emerging work. And with natural language as the input modality, we can always replace the current AudioLDM model with better future models. In this case, the SonifyAR system leaves open ports for future iterations and improvements.
\subsection{Limitations and Failure Cases}
While SonifyAR was positively evaluated in our user study and demonstrated broad application potential, we observed limitations in the system adaptability and encountered some failure cases.

Firstly, since the sound acquisition pipeline relies on an LLM as both the context collector and overall controller, LLM errors can lead to failures. For example, the LLM can hallucinate about the context information and prompt sound generation model with inaccruate text input (\textit{e.g.,} LLM provides prompt ``sliding plastic on carpet'' when the provided context information is the virtual model, a metal toy robot, slides on wooden floor) that lead to non-matching outputs.

Secondly, although AudioLDM~\cite{liu2023audioldm} is a state-of-the-art text-to-audio generator and performs well in most cases, it can sometimes produce subpar outputs that are noisy or do not match prompts. Based on our experiences, such performance inconsistency can be hard to predict thus cannot be completely solved by prompt engineering.

Lastly, some AR interactions can be highly time-sensitive (\textit{e.g.,} objects colliding, playing animations that show clear action sequences) and require sound assets to be precisely synchronized with visual content. \rev{Currently, we use two methods to synchronize sounds with AR events. First, we provide example sound assets for AR interactions of \textit{``Colliding'', ``Tapping''} and \textit{``'Sliding'}. By conducting sound style transfer using well-timed example sounds as inputs, the output sounds can be formatted to be in sync with the AR interaction. Secondly, we bind colliders to 3D model skeleton joints, ensuring that the generated sound is played only when a collision is detected between an animated model and real-world surfaces. However, these synchronization method do not apply to 3D assets without skeletons or sounds unrelated to collisions, such as dinosaur roaring and mechanical arm unfolding. Fully addressing this 3D animation-to-audio synchronization challenge is beyond our current scope.}

%Failure cases can happen in context-to-text (capturing wrong keywords),in sound retrieval (failed to find related sound, sound too long), in sound generation (low quality, non-matching sound), sound not matching rhythm.

%\subsection{Expansion of Framework}
%Currently, we implement the automated AR sound acquisition with six example AR event types, as listed in \autoref{sec:implementation}. Future work should explore other event types, such as \textit{``Spin''}, \textit{``Emphasize''}, \textit{``Expand''}, or \textit{``Jump''}---all which can easily be incorporated into the text template of SonifyAR.

%For wider applicability that reflects the real-world intricacies of AR content creation, we recognize the need to extend its capabilities to accommodate a wider spectrum of AR events and interactions.

%We expect the framework being able to expand to more applications and event types since text is flexible enough. We tried it out in application 3, 4 ,and 5.

\subsection{Future Work}
We plan to address the following improvements for future work. 

First, our current SonifyAR app is a simplified tool that exclusively supports the authoring of AR sound and uses pre-crafted AR assets as input. For wider applicability that reflects the real-world needs of AR content creation, we recognize the need to incorporate the AR sound authoring pipeline into the entire AR content creation process. Our future goals include integrating the SonifyAR framework with established tools like Reality Composer and Adobe Aero, \rev{ and also building a Unity SDK to enable usage of our system among professional AR developers. We will also expand the list of AR event types supported by SonifyAR, adding categories such as \textit{``Spin''}, \textit{``Emphasize''}, \textit{``Expand''}, or \textit{``Jump''}---all of which can easily be incorporated into the text template of SonifyAR.}

Secondly, SonifyAR adopts a straightforward process of using a text template to compile context information collected from multiple sources, including virtual object semantics and real world semantics. Currently, we use one material segmentation model to perceive real-world context information. In future developments, we aspire to incorporate more sophisticated computer vision models or multi-modal foundational models, for deeper understanding of both virtual objects and real world environment, enhancing the context acquisition process with greater precision and broader adaptability.
\rev{
Thirdly, SonifyAR has some basic error-handling features, such as the automatic removal of failed retrieval results and editing options for subpar sound assets (\Cref{subsec:UI}). Additionally,  users can correct recognition errors, such as incorrectly recognized materials, by specifying the correct material with text. In future versions of SonifyAR, we plan to incorporate more advanced error-handling mechanisms, such as deploying a sound quality inference model to validate the output and automatically re-prompt the sound generation model when subpar sounds are generated.}

Lastly, the current sound generation output of SonifyAR presents opportunities for further refinement. Recognizing the rapid advancements in AI, our system has been designed as a modular framework, ensuring that as more advanced models emerge, they can be seamlessly integrated. Every model (\textit{e.g.}, LLM, CV, Sound Generation) within SonifyAR is replaceable, allowing SonifyAR to easily adapt and stay at the forefront of innovation in this field.

\section{Conclusion}
In this paper, we present SonifyAR, a context-aware sound authoring system designed for sonifying AR events. Our work implements a custom PbD authoring pipeline which enables sound asset acquisition using context information and AI. With SonifyAR, users demonstrate AR interactions and have AR sound assets automatically acquired. Our studies validate the usability and overall performance of our system. Our five application examples further supports the potential of our approach. This research advances literature in AR sound authoring, while also opening up new research avenues for the application of LLM and generative models in AR systems.

%Our work offers three main contributions. First, we extend the boundaries of AR sound interaction by introducing a comprehensive AR sound interaction space, emphasizing the significance of user/virtuality-reality interaction sounds. Second, we propose a sound authoring pipeline that takes AR event contextual information and acquire corresponding sound with generative models. Third, we implemented a PbD sound authoring interface and tested it out in an actual application. Our usability study and expert interview verifies our contribution and provide insights that could further broaden the research scope for the realm of AR sound authoring.

\begin{acks}
This research was mainly conducted during an internship at Adobe Research and also supported by NSF award \#1763199. We thank Dr. Cuong Nguyen for his suggestions to our research and our user study participants for their participation. 
\end{acks}

\bibliographystyle{ACM-Reference-Format}
\bibliography{ref}


\begin{thebibliography}{66}

%%% the Web address.

\ifx \showCODEN    \undefined \def \showCODEN     #1{\unskip}     \fi
\ifx \showDOI      \undefined \def \showDOI       #1{#1}\fi
\ifx \showISBNx    \undefined \def \showISBNx     #1{\unskip}     \fi
\ifx \showISBNxiii \undefined \def \showISBNxiii  #1{\unskip}     \fi
\ifx \showISSN     \undefined \def \showISSN      #1{\unskip}     \fi
\ifx \showLCCN     \undefined \def \showLCCN      #1{\unskip}     \fi
\ifx \shownote     \undefined \def \shownote      #1{#1}          \fi
\ifx \showarticletitle \undefined \def \showarticletitle #1{#1}   \fi
\ifx \showURL      \undefined \def \showURL       {\relax}        \fi
% invisible to TeX
\providecommand\bibfield[2]{#2}
\providecommand\bibinfo[2]{#2}
\providecommand\natexlab[1]{#1}
\providecommand\showeprint[2][]{arXiv:#2}

\bibitem[arv({[n.\,d.]})]%
        {arvid}
 \bibinfo{year}{[n.\,d.]}\natexlab{}.
\newblock \bibinfo{booktitle}{\emph{{ARVid - Augmented Reality}}}.
\newblock
\urldef\tempurl%
\url{https://apps.apple.com/us/app/arvid-augmented-reality/id1276546297}
\showURL{%
\tempurl}
\newblock
\shownote{Accessed on September 24, 2023}.


\bibitem[fre({[n.\,d.]})]%
        {freesound}
 \bibinfo{year}{[n.\,d.]}\natexlab{}.
\newblock \bibinfo{booktitle}{\emph{{Freesound}}}.
\newblock
\urldef\tempurl%
\url{https://freesound.org/}
\showURL{%
\tempurl}
\newblock
\shownote{Accessed on September 24, 2023}.


\bibitem[hal({[n.\,d.]})]%
        {haloar}
 \bibinfo{year}{[n.\,d.]}\natexlab{}.
\newblock \bibinfo{booktitle}{\emph{{Halo AR}}}.
\newblock
\urldef\tempurl%
\url{https://haloar.app/}
\showURL{%
\tempurl}
\newblock
\shownote{Accessed on September 24, 2023}.


\bibitem[Adobe(2023b)]%
        {adobe-audition}
\bibfield{author}{\bibinfo{person}{Adobe}.} \bibinfo{year}{2023}\natexlab{b}.
\newblock \bibinfo{booktitle}{\emph{Adobe Audition Sound Effects Download Page}}.
\newblock
\urldef\tempurl%
\url{https://www.adobe.com/products/audition/offers/AdobeAuditionDLCSFX.html}
\showURL{%
\tempurl}
\newblock
\shownote{Accessed on Date of Access}.


\bibitem[Adobe(2023a)]%
        {adobeaero}
\bibfield{author}{\bibinfo{person}{Adobe}.} \bibinfo{year}{Accessed September 11, 2023}\natexlab{a}.
\newblock \bibinfo{title}{Adobe Aero}.
\newblock \bibinfo{howpublished}{\url{https://www.adobe.com/products/aero.html}}.
\newblock


\bibitem[Apple(2023a)]%
        {apple-arkit-doc}
\bibfield{author}{\bibinfo{person}{Apple}.} \bibinfo{year}{2023}\natexlab{a}.
\newblock \bibinfo{booktitle}{\emph{Apple ARKit Documentation}}.
\newblock
\urldef\tempurl%
\url{https://developer.apple.com/documentation/arkit/}
\showURL{%
\tempurl}
\newblock
\shownote{Accessed on Oct 9th, 2023}.


\bibitem[Apple(2023b)]%
        {apple-arkit-plane-tracking}
\bibfield{author}{\bibinfo{person}{Apple}.} \bibinfo{year}{2023}\natexlab{b}.
\newblock \bibinfo{booktitle}{\emph{ARKit - Tracking and Visualizing Planes}}.
\newblock
\urldef\tempurl%
\url{https://developer.apple.com/documentation/arkit/arkit_in_ios/content_anchors/tracking_and_visualizing_planes}
\showURL{%
\tempurl}
\newblock
\shownote{Accessed on Oct 9th, 2023}.


\bibitem[Apple(2023d)]%
        {apple-scenekit-physics-simulation}
\bibfield{author}{\bibinfo{person}{Apple}.} \bibinfo{year}{2023}\natexlab{d}.
\newblock \bibinfo{booktitle}{\emph{SceneKit - Physics Simulation}}.
\newblock
\urldef\tempurl%
\url{https://developer.apple.com/documentation/scenekit/physics_simulation}
\showURL{%
\tempurl}
\newblock
\shownote{Accessed on Oct 9th, 2023}.


\bibitem[Apple(2023c)]%
        {realitycomposer}
\bibfield{author}{\bibinfo{person}{Apple}.} \bibinfo{year}{Accessed September 11, 2023}\natexlab{c}.
\newblock \bibinfo{title}{Reality Composer}.
\newblock \bibinfo{howpublished}{\url{https://apps.apple.com/us/app/reality-composer/id1462358802}}.
\newblock


\bibitem[Chen et~al\mbox{.}(2023)]%
        {chen2023vast}
\bibfield{author}{\bibinfo{person}{Sihan Chen}, \bibinfo{person}{Handong Li}, \bibinfo{person}{Qunbo Wang}, \bibinfo{person}{Zijia Zhao}, \bibinfo{person}{Mingzhen Sun}, \bibinfo{person}{Xinxin Zhu}, {and} \bibinfo{person}{Jing Liu}.} \bibinfo{year}{2023}\natexlab{}.
\newblock \bibinfo{title}{VAST: A Vision-Audio-Subtitle-Text Omni-Modality Foundation Model and Dataset}.
\newblock
\newblock
\showeprint[arxiv]{2305.18500}~[cs.CV]


\bibitem[Chulpongsatorn et~al\mbox{.}(2023)]%
        {Chulpongsatorn_AugmentedMath_UIST2023}
\bibfield{author}{\bibinfo{person}{Neil Chulpongsatorn}, \bibinfo{person}{Mille~Skovhus Lunding}, \bibinfo{person}{Nishan Soni}, {and} \bibinfo{person}{Ryo Suzuki}.} \bibinfo{year}{2023}\natexlab{}.
\newblock \showarticletitle{Augmented Math: Authoring AR-Based Explorable Explanations by Augmenting Static Math Textbooks}. In \bibinfo{booktitle}{\emph{Proceedings of the 36th Annual ACM Symposium on User Interface Software and Technology}} (San Francisco, CA, USA) \emph{(\bibinfo{series}{UIST '23})}. \bibinfo{publisher}{Association for Computing Machinery}, \bibinfo{address}{New York, NY, USA}, Article \bibinfo{articleno}{92}, \bibinfo{numpages}{16}~pages.
\newblock
\showISBNx{9798400701320}
\urldef\tempurl%
\url{https://doi.org/10.1145/3586183.3606827}
\showDOI{\tempurl}


\bibitem[Coughlan and Miele(2017)]%
        {coughlan2017ar4vi}
\bibfield{author}{\bibinfo{person}{James~M Coughlan} {and} \bibinfo{person}{Joshua Miele}.} \bibinfo{year}{2017}\natexlab{}.
\newblock \showarticletitle{AR4VI: AR as an accessibility tool for people with visual impairments}. In \bibinfo{booktitle}{\emph{2017 IEEE International Symposium on Mixed and Augmented Reality (ISMAR-Adjunct)}}. IEEE, \bibinfo{pages}{288--292}.
\newblock


\bibitem[Dam et~al\mbox{.}(2024)]%
        {dam2024taxonomy}
\bibfield{author}{\bibinfo{person}{Abhraneil Dam}, \bibinfo{person}{Arsh Siddiqui}, \bibinfo{person}{Charles Leclercq}, {and} \bibinfo{person}{Myounghoon Jeon}.} \bibinfo{year}{2024}\natexlab{}.
\newblock \showarticletitle{Taxonomy and definition of audio augmented reality (AAR): A grounded theory study}.
\newblock \bibinfo{journal}{\emph{International Journal of Human-Computer Studies}}  \bibinfo{volume}{182} (\bibinfo{year}{2024}), \bibinfo{pages}{103179}.
\newblock


\bibitem[Dhariwal et~al\mbox{.}(2020)]%
        {dhariwal2020jukebox}
\bibfield{author}{\bibinfo{person}{Prafulla Dhariwal}, \bibinfo{person}{Heewoo Jun}, \bibinfo{person}{Christine Payne}, \bibinfo{person}{Jong~Wook Kim}, \bibinfo{person}{Alec Radford}, {and} \bibinfo{person}{Ilya Sutskever}.} \bibinfo{year}{2020}\natexlab{}.
\newblock \showarticletitle{Jukebox: A generative model for music}.
\newblock \bibinfo{journal}{\emph{arXiv preprint arXiv:2005.00341}} (\bibinfo{year}{2020}).
\newblock


\bibitem[Diaz et~al\mbox{.}(2022)]%
        {diaz_rigid-body_2022}
\bibfield{author}{\bibinfo{person}{Rodrigo Diaz}, \bibinfo{person}{Ben Hayes}, \bibinfo{person}{Charalampos Saitis}, \bibinfo{person}{György Fazekas}, {and} \bibinfo{person}{Mark Sandler}.} \bibinfo{year}{2022}\natexlab{}.
\newblock \bibinfo{title}{Rigid-{Body} {Sound} {Synthesis} with {Differentiable} {Modal} {Resonators}}.
\newblock
\newblock
\urldef\tempurl%
\url{http://arxiv.org/abs/2210.15306}
\showURL{%
\tempurl}
\newblock
\shownote{arXiv:2210.15306 [cs, eess]}.


\bibitem[Engel et~al\mbox{.}(2019)]%
        {engel2019gansynth}
\bibfield{author}{\bibinfo{person}{Jesse Engel}, \bibinfo{person}{Kumar~Krishna Agrawal}, \bibinfo{person}{Shuo Chen}, \bibinfo{person}{Ishaan Gulrajani}, \bibinfo{person}{Chris Donahue}, {and} \bibinfo{person}{Adam Roberts}.} \bibinfo{year}{2019}\natexlab{}.
\newblock \showarticletitle{Gansynth: Adversarial neural audio synthesis}.
\newblock \bibinfo{journal}{\emph{arXiv preprint arXiv:1902.08710}} (\bibinfo{year}{2019}).
\newblock


\bibitem[Filus and Rambli(2012)]%
        {filus2012using}
\bibfield{author}{\bibinfo{person}{Mohd Ihsan Alimi~Mohd Filus} {and} \bibinfo{person}{Dayang Rohaya~Awang Rambli}.} \bibinfo{year}{2012}\natexlab{}.
\newblock \showarticletitle{Using non-speech sound as acoustic modality in Augmented Reality environment}. In \bibinfo{booktitle}{\emph{2012 International Symposium on Computer Applications and Industrial Electronics (ISCAIE)}}. IEEE, \bibinfo{pages}{79--82}.
\newblock


\bibitem[Games(2023)]%
        {unreal-engine-website}
\bibfield{author}{\bibinfo{person}{Epic Games}.} \bibinfo{year}{2023}\natexlab{}.
\newblock \bibinfo{booktitle}{\emph{Unreal Engine}}.
\newblock
\urldef\tempurl%
\url{https://www.unrealengine.com/en-US}
\showURL{%
\tempurl}
\newblock
\shownote{Accessed on Oct 9th, 2023}.


\bibitem[Ghose and Prevost(2022)]%
        {ghose2022foleygan}
\bibfield{author}{\bibinfo{person}{Sanchita Ghose} {and} \bibinfo{person}{John~J Prevost}.} \bibinfo{year}{2022}\natexlab{}.
\newblock \showarticletitle{Foleygan: Visually guided generative adversarial network-based synchronous sound generation in silent videos}.
\newblock \bibinfo{journal}{\emph{IEEE Transactions on Multimedia}} (\bibinfo{year}{2022}).
\newblock


\bibitem[Han et~al\mbox{.}(2020)]%
        {han2020live}
\bibfield{author}{\bibinfo{person}{Lei Han}, \bibinfo{person}{Tian Zheng}, \bibinfo{person}{Yinheng Zhu}, \bibinfo{person}{Lan Xu}, {and} \bibinfo{person}{Lu Fang}.} \bibinfo{year}{2020}\natexlab{}.
\newblock \showarticletitle{Live semantic 3d perception for immersive augmented reality}.
\newblock \bibinfo{journal}{\emph{IEEE transactions on visualization and computer graphics}} \bibinfo{volume}{26}, \bibinfo{number}{5} (\bibinfo{year}{2020}), \bibinfo{pages}{2012--2022}.
\newblock


\bibitem[Irwansyah et~al\mbox{.}(2018)]%
        {irwansyah2018augmented}
\bibfield{author}{\bibinfo{person}{Ferli~Septi Irwansyah}, \bibinfo{person}{YM Yusuf}, \bibinfo{person}{Ida Farida}, {and} \bibinfo{person}{Muhammad~Ali Ramdhani}.} \bibinfo{year}{2018}\natexlab{}.
\newblock \showarticletitle{Augmented reality (AR) technology on the android operating system in chemistry learning}. In \bibinfo{booktitle}{\emph{IOP conference series: Materials science and engineering}}, Vol.~\bibinfo{volume}{288}. IOP Publishing, \bibinfo{pages}{012068}.
\newblock


\bibitem[Jain et~al\mbox{.}(2021)]%
        {jain_taxonomy_2021}
\bibfield{author}{\bibinfo{person}{Dhruv Jain}, \bibinfo{person}{Sasa Junuzovic}, \bibinfo{person}{Eyal Ofek}, \bibinfo{person}{Mike Sinclair}, \bibinfo{person}{John R.~Porter}, \bibinfo{person}{Chris Yoon}, \bibinfo{person}{Swetha Machanavajhala}, {and} \bibinfo{person}{Meredith Ringel~Morris}.} \bibinfo{year}{2021}\natexlab{}.
\newblock \showarticletitle{A {Taxonomy} of {Sounds} in {Virtual} {Reality}}. In \bibinfo{booktitle}{\emph{Proceedings of the 2021 {International} {Conference} on {Multimodal} {Interaction}}}. \bibinfo{publisher}{ACM}, \bibinfo{address}{Montréal QC Canada}, \bibinfo{pages}{80--91}.
\newblock
\showISBNx{978-1-4503-8481-0}
\urldef\tempurl%
\url{https://doi.org/10.1145/3462244.3479946}
\showDOI{\tempurl}


\bibitem[Jin et~al\mbox{.}(2020)]%
        {jin_deep-modal_2020}
\bibfield{author}{\bibinfo{person}{Xutong Jin}, \bibinfo{person}{Sheng Li}, \bibinfo{person}{Tianshu Qu}, \bibinfo{person}{Dinesh Manocha}, {and} \bibinfo{person}{Guoping Wang}.} \bibinfo{year}{2020}\natexlab{}.
\newblock \showarticletitle{Deep-{Modal}: {Real}-{Time} {Impact} {Sound} {Synthesis} for {Arbitrary} {Shapes}}. In \bibinfo{booktitle}{\emph{Proceedings of the 28th {ACM} {International} {Conference} on {Multimedia}}} \emph{(\bibinfo{series}{{MM} '20})}. \bibinfo{publisher}{Association for Computing Machinery}, \bibinfo{address}{New York, NY, USA}, \bibinfo{pages}{1171--1179}.
\newblock
\showISBNx{978-1-4503-7988-5}
\urldef\tempurl%
\url{https://doi.org/10.1145/3394171.3413572}
\showDOI{\tempurl}


\bibitem[Jin et~al\mbox{.}(2022)]%
        {jin_neuralsound_2022}
\bibfield{author}{\bibinfo{person}{Xutong Jin}, \bibinfo{person}{Sheng Li}, \bibinfo{person}{Guoping Wang}, {and} \bibinfo{person}{Dinesh Manocha}.} \bibinfo{year}{2022}\natexlab{}.
\newblock \showarticletitle{{NeuralSound}: learning-based modal sound synthesis with acoustic transfer}.
\newblock \bibinfo{journal}{\emph{ACM Transactions on Graphics}} \bibinfo{volume}{41}, \bibinfo{number}{4} (\bibinfo{date}{July} \bibinfo{year}{2022}), \bibinfo{pages}{1--15}.
\newblock
\showISSN{0730-0301, 1557-7368}
\urldef\tempurl%
\url{https://doi.org/10.1145/3528223.3530184}
\showDOI{\tempurl}


\bibitem[Kang et~al\mbox{.}(2019)]%
        {Kang_PrototypAR_IDC2019}
\bibfield{author}{\bibinfo{person}{Seokbin Kang}, \bibinfo{person}{Leyla Norooz}, \bibinfo{person}{Elizabeth Bonsignore}, \bibinfo{person}{Virginia Byrne}, \bibinfo{person}{Tamara Clegg}, {and} \bibinfo{person}{Jon~E. Froehlich}.} \bibinfo{year}{2019}\natexlab{}.
\newblock \showarticletitle{PrototypAR: Prototyping and Simulating Complex Systems with Paper Craft and Augmented Reality}. In \bibinfo{booktitle}{\emph{Proceedings of the 18th ACM International Conference on Interaction Design and Children}} (Boise, ID, USA) \emph{(\bibinfo{series}{IDC '19})}. \bibinfo{publisher}{Association for Computing Machinery}, \bibinfo{address}{New York, NY, USA}, \bibinfo{pages}{253–266}.
\newblock
\showISBNx{9781450366908}
\urldef\tempurl%
\url{https://doi.org/10.1145/3311927.3323135}
\showDOI{\tempurl}


\bibitem[Kang et~al\mbox{.}(2016)]%
        {Kang_SharedPhys_IDC2016}
\bibfield{author}{\bibinfo{person}{Seokbin Kang}, \bibinfo{person}{Leyla Norooz}, \bibinfo{person}{Vanessa Oguamanam}, \bibinfo{person}{Angelisa~C. Plane}, \bibinfo{person}{Tamara~L. Clegg}, {and} \bibinfo{person}{Jon~E. Froehlich}.} \bibinfo{year}{2016}\natexlab{}.
\newblock \showarticletitle{SharedPhys: Live Physiological Sensing, Whole-Body Interaction, and Large-Screen Visualizations to Support Shared Inquiry Experiences}. In \bibinfo{booktitle}{\emph{Proceedings of the The 15th International Conference on Interaction Design and Children}} (Manchester, United Kingdom) \emph{(\bibinfo{series}{IDC '16})}. \bibinfo{publisher}{Association for Computing Machinery}, \bibinfo{address}{New York, NY, USA}, \bibinfo{pages}{275–287}.
\newblock
\showISBNx{9781450343138}
\urldef\tempurl%
\url{https://doi.org/10.1145/2930674.2930710}
\showDOI{\tempurl}


\bibitem[Kang et~al\mbox{.}(2020)]%
        {Kang_ARMath_CHI2020}
\bibfield{author}{\bibinfo{person}{Seokbin Kang}, \bibinfo{person}{Ekta Shokeen}, \bibinfo{person}{Virginia~L. Byrne}, \bibinfo{person}{Leyla Norooz}, \bibinfo{person}{Elizabeth Bonsignore}, \bibinfo{person}{Caro Williams-Pierce}, {and} \bibinfo{person}{Jon~E. Froehlich}.} \bibinfo{year}{2020}\natexlab{}.
\newblock \showarticletitle{ARMath: Augmenting Everyday Life with Math Learning}. In \bibinfo{booktitle}{\emph{Proceedings of the 2020 CHI Conference on Human Factors in Computing Systems}} (Honolulu, HI, USA) \emph{(\bibinfo{series}{CHI '20})}. \bibinfo{publisher}{Association for Computing Machinery}, \bibinfo{address}{New York, NY, USA}, \bibinfo{pages}{1–15}.
\newblock
\showISBNx{9781450367080}
\urldef\tempurl%
\url{https://doi.org/10.1145/3313831.3376252}
\showDOI{\tempurl}


\bibitem[Koepke et~al\mbox{.}(2022)]%
        {koepke2022audio}
\bibfield{author}{\bibinfo{person}{A~Sophia Koepke}, \bibinfo{person}{Andreea-Maria Oncescu}, \bibinfo{person}{Jo{\~a}o~F Henriques}, \bibinfo{person}{Zeynep Akata}, {and} \bibinfo{person}{Samuel Albanie}.} \bibinfo{year}{2022}\natexlab{}.
\newblock \showarticletitle{Audio retrieval with natural language queries: A benchmark study}.
\newblock \bibinfo{journal}{\emph{IEEE Transactions on Multimedia}}  \bibinfo{volume}{25} (\bibinfo{year}{2022}), \bibinfo{pages}{2675--2685}.
\newblock


\bibitem[Kong et~al\mbox{.}(2020)]%
        {kong2020diffwave}
\bibfield{author}{\bibinfo{person}{Zhifeng Kong}, \bibinfo{person}{Wei Ping}, \bibinfo{person}{Jiaji Huang}, \bibinfo{person}{Kexin Zhao}, {and} \bibinfo{person}{Bryan Catanzaro}.} \bibinfo{year}{2020}\natexlab{}.
\newblock \showarticletitle{Diffwave: A versatile diffusion model for audio synthesis}.
\newblock \bibinfo{journal}{\emph{arXiv preprint arXiv:2009.09761}} (\bibinfo{year}{2020}).
\newblock


\bibitem[Krings et~al\mbox{.}(2020)]%
        {krings_development_2020}
\bibfield{author}{\bibinfo{person}{Sarah Krings}, \bibinfo{person}{Enes Yigitbas}, \bibinfo{person}{Ivan Jovanovikj}, \bibinfo{person}{Stefan Sauer}, {and} \bibinfo{person}{Gregor Engels}.} \bibinfo{year}{2020}\natexlab{}.
\newblock \showarticletitle{Development framework for context-aware augmented reality applications}. In \bibinfo{booktitle}{\emph{Companion {Proceedings} of the 12th {ACM} {SIGCHI} {Symposium} on {Engineering} {Interactive} {Computing} {Systems}}} \emph{(\bibinfo{series}{{EICS} '20 {Companion}})}. \bibinfo{publisher}{Association for Computing Machinery}, \bibinfo{address}{New York, NY, USA}, \bibinfo{pages}{1--6}.
\newblock
\showISBNx{978-1-4503-7984-7}
\urldef\tempurl%
\url{https://doi.org/10.1145/3393672.3398640}
\showDOI{\tempurl}


\bibitem[Kumar et~al\mbox{.}(2019)]%
        {kumar2019melgan}
\bibfield{author}{\bibinfo{person}{Kundan Kumar}, \bibinfo{person}{Rithesh Kumar}, \bibinfo{person}{Thibault De~Boissiere}, \bibinfo{person}{Lucas Gestin}, \bibinfo{person}{Wei~Zhen Teoh}, \bibinfo{person}{Jose Sotelo}, \bibinfo{person}{Alexandre De~Brebisson}, \bibinfo{person}{Yoshua Bengio}, {and} \bibinfo{person}{Aaron~C Courville}.} \bibinfo{year}{2019}\natexlab{}.
\newblock \showarticletitle{Melgan: Generative adversarial networks for conditional waveform synthesis}.
\newblock \bibinfo{journal}{\emph{Advances in neural information processing systems}}  \bibinfo{volume}{32} (\bibinfo{year}{2019}).
\newblock


\bibitem[Lang et~al\mbox{.}(2019)]%
        {lang2019virtual}
\bibfield{author}{\bibinfo{person}{Yining Lang}, \bibinfo{person}{Wei Liang}, {and} \bibinfo{person}{Lap-Fai Yu}.} \bibinfo{year}{2019}\natexlab{}.
\newblock \showarticletitle{Virtual agent positioning driven by scene semantics in mixed reality}. In \bibinfo{booktitle}{\emph{2019 IEEE Conference on Virtual Reality and 3D User Interfaces (VR)}}. IEEE, \bibinfo{pages}{767--775}.
\newblock


\bibitem[Lau and Weld(1998)]%
        {10.1145/291080.291104}
\bibfield{author}{\bibinfo{person}{Tessa~A. Lau} {and} \bibinfo{person}{Daniel~S. Weld}.} \bibinfo{year}{1998}\natexlab{}.
\newblock \showarticletitle{Programming by Demonstration: An Inductive Learning Formulation}. In \bibinfo{booktitle}{\emph{Proceedings of the 4th International Conference on Intelligent User Interfaces}} (Los Angeles, California, USA) \emph{(\bibinfo{series}{IUI '99})}. \bibinfo{publisher}{Association for Computing Machinery}, \bibinfo{address}{New York, NY, USA}, \bibinfo{pages}{145–152}.
\newblock
\showISBNx{1581130988}
\urldef\tempurl%
\url{https://doi.org/10.1145/291080.291104}
\showDOI{\tempurl}


\bibitem[Liang et~al\mbox{.}(2021)]%
        {liang2021scene}
\bibfield{author}{\bibinfo{person}{Wei Liang}, \bibinfo{person}{Xinzhe Yu}, \bibinfo{person}{Rawan Alghofaili}, \bibinfo{person}{Yining Lang}, {and} \bibinfo{person}{Lap-Fai Yu}.} \bibinfo{year}{2021}\natexlab{}.
\newblock \showarticletitle{Scene-aware behavior synthesis for virtual pets in mixed reality}. In \bibinfo{booktitle}{\emph{Proceedings of the 2021 CHI Conference on Human Factors in Computing Systems}}. \bibinfo{pages}{1--12}.
\newblock


\bibitem[Lin et~al\mbox{.}(2023)]%
        {lin2023soundify}
\bibfield{author}{\bibinfo{person}{David Chuan-En Lin}, \bibinfo{person}{Anastasis Germanidis}, \bibinfo{person}{Crist{\'o}bal Valenzuela}, \bibinfo{person}{Yining Shi}, {and} \bibinfo{person}{Nikolas Martelaro}.} \bibinfo{year}{2023}\natexlab{}.
\newblock \showarticletitle{Soundify: Matching sound effects to video}. In \bibinfo{booktitle}{\emph{Proceedings of the 36th Annual ACM Symposium on User Interface Software and Technology}}. \bibinfo{pages}{1--13}.
\newblock


\bibitem[Lindlbauer et~al\mbox{.}(2019)]%
        {lindlbauer2019context}
\bibfield{author}{\bibinfo{person}{David Lindlbauer}, \bibinfo{person}{Anna~Maria Feit}, {and} \bibinfo{person}{Otmar Hilliges}.} \bibinfo{year}{2019}\natexlab{}.
\newblock \showarticletitle{Context-aware online adaptation of mixed reality interfaces}. In \bibinfo{booktitle}{\emph{Proceedings of the 32nd annual ACM symposium on user interface software and technology}}. \bibinfo{pages}{147--160}.
\newblock


\bibitem[Liu et~al\mbox{.}(2023)]%
        {liu2023audioldm}
\bibfield{author}{\bibinfo{person}{Haohe Liu}, \bibinfo{person}{Zehua Chen}, \bibinfo{person}{Yi Yuan}, \bibinfo{person}{Xinhao Mei}, \bibinfo{person}{Xubo Liu}, \bibinfo{person}{Danilo Mandic}, \bibinfo{person}{Wenwu Wang}, {and} \bibinfo{person}{Mark~D Plumbley}.} \bibinfo{year}{2023}\natexlab{}.
\newblock \showarticletitle{Audioldm: Text-to-audio generation with latent diffusion models}.
\newblock \bibinfo{journal}{\emph{arXiv preprint arXiv:2301.12503}} (\bibinfo{year}{2023}).
\newblock


\bibitem[Liu and Manocha(2021)]%
        {liu_sound_2021}
\bibfield{author}{\bibinfo{person}{Shiguang Liu} {and} \bibinfo{person}{Dinesh Manocha}.} \bibinfo{year}{2021}\natexlab{}.
\newblock \bibinfo{title}{Sound {Synthesis}, {Propagation}, and {Rendering}: {A} {Survey}}.
\newblock
\newblock
\urldef\tempurl%
\url{http://arxiv.org/abs/2011.05538}
\showURL{%
\tempurl}
\newblock
\shownote{arXiv:2011.05538 [cs]}.


\bibitem[Monteiro et~al\mbox{.}(2023)]%
        {monteiro2023teachable}
\bibfield{author}{\bibinfo{person}{Kyzyl Monteiro}, \bibinfo{person}{Ritik Vatsal}, \bibinfo{person}{Neil Chulpongsatorn}, \bibinfo{person}{Aman Parnami}, {and} \bibinfo{person}{Ryo Suzuki}.} \bibinfo{year}{2023}\natexlab{}.
\newblock \showarticletitle{Teachable Reality: Prototyping Tangible Augmented Reality with Everyday Objects by Leveraging Interactive Machine Teaching}. In \bibinfo{booktitle}{\emph{Proceedings of the 2023 CHI Conference on Human Factors in Computing Systems}}. \bibinfo{pages}{1--15}.
\newblock


\bibitem[Nebeling and Speicher(2018)]%
        {nebeling_trouble_2018}
\bibfield{author}{\bibinfo{person}{Michael Nebeling} {and} \bibinfo{person}{Maximilian Speicher}.} \bibinfo{year}{2018}\natexlab{}.
\newblock \showarticletitle{The {Trouble} with {Augmented} {Reality}/{Virtual} {Reality} {Authoring} {Tools}}. In \bibinfo{booktitle}{\emph{2018 {IEEE} {International} {Symposium} on {Mixed} and {Augmented} {Reality} {Adjunct} ({ISMAR}-{Adjunct})}}. \bibinfo{publisher}{IEEE}, \bibinfo{address}{Munich, Germany}, \bibinfo{pages}{333--337}.
\newblock
\showISBNx{978-1-5386-7592-2}
\urldef\tempurl%
\url{https://doi.org/10.1109/ISMAR-Adjunct.2018.00098}
\showDOI{\tempurl}


\bibitem[Niantic(2024)]%
        {pokemon_go}
\bibfield{author}{\bibinfo{person}{Inc. Niantic}.} \bibinfo{year}{2024}\natexlab{}.
\newblock \bibinfo{title}{Pokémon GO}.
\newblock \bibinfo{howpublished}{\url{https://pokemongolive.com/?hl=en}}.
\newblock
\newblock
\shownote{Accessed: 2024-07-22}.


\bibitem[OpenAI(2024)]%
        {openai2024gpt4}
\bibfield{author}{\bibinfo{person}{OpenAI}.} \bibinfo{year}{2024}\natexlab{}.
\newblock \bibinfo{title}{GPT-4 Technical Report}.
\newblock
\newblock
\showeprint[arxiv]{2303.08774}~[cs.CL]


\bibitem[Qian et~al\mbox{.}(2022)]%
        {qian2022scalar}
\bibfield{author}{\bibinfo{person}{Xun Qian}, \bibinfo{person}{Fengming He}, \bibinfo{person}{Xiyun Hu}, \bibinfo{person}{Tianyi Wang}, \bibinfo{person}{Ananya Ipsita}, {and} \bibinfo{person}{Karthik Ramani}.} \bibinfo{year}{2022}\natexlab{}.
\newblock \showarticletitle{Scalar: Authoring semantically adaptive augmented reality experiences in virtual reality}. In \bibinfo{booktitle}{\emph{Proceedings of the 2022 CHI Conference on Human Factors in Computing Systems}}. \bibinfo{pages}{1--18}.
\newblock


\bibitem[Radu and Schneider(2019)]%
        {radu2019can}
\bibfield{author}{\bibinfo{person}{Iulian Radu} {and} \bibinfo{person}{Bertrand Schneider}.} \bibinfo{year}{2019}\natexlab{}.
\newblock \showarticletitle{What can we learn from augmented reality (AR)? Benefits and drawbacks of AR for inquiry-based learning of physics}. In \bibinfo{booktitle}{\emph{Proceedings of the 2019 CHI conference on human factors in computing systems}}. \bibinfo{pages}{1--12}.
\newblock


\bibitem[Raghuvanshi and Lin(2006)]%
        {raghuvanshi_interactive_2006}
\bibfield{author}{\bibinfo{person}{Nikunj Raghuvanshi} {and} \bibinfo{person}{Ming~C. Lin}.} \bibinfo{year}{2006}\natexlab{}.
\newblock \showarticletitle{Interactive sound synthesis for large scale environments}. In \bibinfo{booktitle}{\emph{Proceedings of the 2006 symposium on {Interactive} {3D} graphics and games}} \emph{(\bibinfo{series}{{I3D} '06})}. \bibinfo{publisher}{Association for Computing Machinery}, \bibinfo{address}{New York, NY, USA}, \bibinfo{pages}{101--108}.
\newblock
\showISBNx{978-1-59593-295-2}
\urldef\tempurl%
\url{https://doi.org/10.1145/1111411.1111429}
\showDOI{\tempurl}


\bibitem[Rakkolainen et~al\mbox{.}(2021)]%
        {rakkolainen2021technologies}
\bibfield{author}{\bibinfo{person}{Ismo Rakkolainen}, \bibinfo{person}{Ahmed Farooq}, \bibinfo{person}{Jari Kangas}, \bibinfo{person}{Jaakko Hakulinen}, \bibinfo{person}{Jussi Rantala}, \bibinfo{person}{Markku Turunen}, {and} \bibinfo{person}{Roope Raisamo}.} \bibinfo{year}{2021}\natexlab{}.
\newblock \showarticletitle{Technologies for multimodal interaction in extended reality—a scoping review}.
\newblock \bibinfo{journal}{\emph{Multimodal Technologies and Interaction}} \bibinfo{volume}{5}, \bibinfo{number}{12} (\bibinfo{year}{2021}), \bibinfo{pages}{81}.
\newblock


\bibitem[Ren et~al\mbox{.}(2013)]%
        {ren_example-guided_2013}
\bibfield{author}{\bibinfo{person}{Zhimin Ren}, \bibinfo{person}{Hengchin Yeh}, {and} \bibinfo{person}{Ming~C. Lin}.} \bibinfo{year}{2013}\natexlab{}.
\newblock \showarticletitle{Example-guided physically based modal sound synthesis}.
\newblock \bibinfo{journal}{\emph{ACM Transactions on Graphics}} \bibinfo{volume}{32}, \bibinfo{number}{1} (\bibinfo{date}{Feb.} \bibinfo{year}{2013}), \bibinfo{pages}{1:1--1:16}.
\newblock
\showISSN{0730-0301}
\urldef\tempurl%
\url{https://doi.org/10.1145/2421636.2421637}
\showDOI{\tempurl}


\bibitem[Ribeiro et~al\mbox{.}(2012a)]%
        {ribeiro2012auditory}
\bibfield{author}{\bibinfo{person}{Flavio Ribeiro}, \bibinfo{person}{Dinei Florencio}, \bibinfo{person}{Philip~A Chou}, {and} \bibinfo{person}{Zhengyou Zhang}.} \bibinfo{year}{2012}\natexlab{a}.
\newblock \showarticletitle{Auditory augmented reality: Object sonification for the visually impaired}. In \bibinfo{booktitle}{\emph{2012 IEEE 14th international workshop on multimedia signal processing (MMSP)}}. IEEE, \bibinfo{pages}{319--324}.
\newblock


\bibitem[Ribeiro et~al\mbox{.}(2012b)]%
        {ribeiro_auditory_2012}
\bibfield{author}{\bibinfo{person}{Flávio Ribeiro}, \bibinfo{person}{Dinei Florêncio}, \bibinfo{person}{Philip~A. Chou}, {and} \bibinfo{person}{Zhengyou Zhang}.} \bibinfo{year}{2012}\natexlab{b}.
\newblock \showarticletitle{Auditory augmented reality: {Object} sonification for the visually impaired}. In \bibinfo{booktitle}{\emph{2012 {IEEE} 14th {International} {Workshop} on {Multimedia} {Signal} {Processing} ({MMSP})}}. \bibinfo{pages}{319--324}.
\newblock
\urldef\tempurl%
\url{https://doi.org/10.1109/MMSP.2012.6343462}
\showDOI{\tempurl}


\bibitem[Roginska and Geluso(2017)]%
        {roginska2017immersive}
\bibfield{author}{\bibinfo{person}{Agnieszka Roginska} {and} \bibinfo{person}{Paul Geluso}.} \bibinfo{year}{2017}\natexlab{}.
\newblock \bibinfo{booktitle}{\emph{Immersive Sound}}.
\newblock \bibinfo{publisher}{Focal Press}.
\newblock


\bibitem[Roodaki et~al\mbox{.}(2017)]%
        {roodaki_sonifeye_2017}
\bibfield{author}{\bibinfo{person}{Hessam Roodaki}, \bibinfo{person}{Navid Navab}, \bibinfo{person}{Abouzar Eslami}, \bibinfo{person}{Christopher Stapleton}, {and} \bibinfo{person}{Nassir Navab}.} \bibinfo{year}{2017}\natexlab{}.
\newblock \showarticletitle{{SonifEye}: {Sonification} of {Visual} {Information} {Using} {Physical} {Modeling} {Sound} {Synthesis}}.
\newblock \bibinfo{journal}{\emph{IEEE Transactions on Visualization and Computer Graphics}} \bibinfo{volume}{23}, \bibinfo{number}{11} (\bibinfo{date}{Nov.} \bibinfo{year}{2017}), \bibinfo{pages}{2366--2371}.
\newblock
\showISSN{1941-0506}
\urldef\tempurl%
\url{https://doi.org/10.1109/TVCG.2017.2734327}
\showDOI{\tempurl}
\newblock
\shownote{Conference Name: IEEE Transactions on Visualization and Computer Graphics}.


\bibitem[Rumi{\'n}ski(2015)]%
        {ruminski2015experimental}
\bibfield{author}{\bibinfo{person}{Dariusz Rumi{\'n}ski}.} \bibinfo{year}{2015}\natexlab{}.
\newblock \showarticletitle{An experimental study of spatial sound usefulness in searching and navigating through AR environments}.
\newblock \bibinfo{journal}{\emph{Virtual Reality}} \bibinfo{volume}{19}, \bibinfo{number}{3-4} (\bibinfo{year}{2015}), \bibinfo{pages}{223--233}.
\newblock


\bibitem[Serafin et~al\mbox{.}(2018)]%
        {serafin_sonic_2018}
\bibfield{author}{\bibinfo{person}{Stefania Serafin}, \bibinfo{person}{Michele Geronazzo}, \bibinfo{person}{Cumhur Erkut}, \bibinfo{person}{Niels~C. Nilsson}, {and} \bibinfo{person}{Rolf Nordahl}.} \bibinfo{year}{2018}\natexlab{}.
\newblock \showarticletitle{Sonic {Interactions} in {Virtual} {Reality}: {State} of the {Art}, {Current} {Challenges}, and {Future} {Directions}}.
\newblock \bibinfo{journal}{\emph{IEEE Computer Graphics and Applications}} \bibinfo{volume}{38}, \bibinfo{number}{2} (\bibinfo{date}{March} \bibinfo{year}{2018}), \bibinfo{pages}{31--43}.
\newblock
\showISSN{1558-1756}
\urldef\tempurl%
\url{https://doi.org/10.1109/MCG.2018.193142628}
\showDOI{\tempurl}
\newblock
\shownote{Conference Name: IEEE Computer Graphics and Applications}.


\bibitem[Sheffer and Adi(2023)]%
        {10096023}
\bibfield{author}{\bibinfo{person}{Roy Sheffer} {and} \bibinfo{person}{Yossi Adi}.} \bibinfo{year}{2023}\natexlab{}.
\newblock \showarticletitle{I Hear Your True Colors: Image Guided Audio Generation}. In \bibinfo{booktitle}{\emph{ICASSP 2023 - 2023 IEEE International Conference on Acoustics, Speech and Signal Processing (ICASSP)}}. \bibinfo{pages}{1--5}.
\newblock
\urldef\tempurl%
\url{https://doi.org/10.1109/ICASSP49357.2023.10096023}
\showDOI{\tempurl}


\bibitem[Shen et~al\mbox{.}(2023)]%
        {shen2023hugginggpt}
\bibfield{author}{\bibinfo{person}{Yongliang Shen}, \bibinfo{person}{Kaitao Song}, \bibinfo{person}{Xu Tan}, \bibinfo{person}{Dongsheng Li}, \bibinfo{person}{Weiming Lu}, {and} \bibinfo{person}{Yueting Zhuang}.} \bibinfo{year}{2023}\natexlab{}.
\newblock \showarticletitle{Hugginggpt: Solving ai tasks with chatgpt and its friends in huggingface}.
\newblock \bibinfo{journal}{\emph{arXiv preprint arXiv:2303.17580}} (\bibinfo{year}{2023}).
\newblock


\bibitem[S{\i}rakaya and Alsancak~S{\i}rakaya(2022)]%
        {sirakaya2022augmented}
\bibfield{author}{\bibinfo{person}{Mustafa S{\i}rakaya} {and} \bibinfo{person}{Didem Alsancak~S{\i}rakaya}.} \bibinfo{year}{2022}\natexlab{}.
\newblock \showarticletitle{Augmented reality in STEM education: A systematic review}.
\newblock \bibinfo{journal}{\emph{Interactive Learning Environments}} \bibinfo{volume}{30}, \bibinfo{number}{8} (\bibinfo{year}{2022}), \bibinfo{pages}{1556--1569}.
\newblock


\bibitem[Suzuki et~al\mbox{.}(2020)]%
        {Suzuki_RealitySketch_UIST2020}
\bibfield{author}{\bibinfo{person}{Ryo Suzuki}, \bibinfo{person}{Rubaiat~Habib Kazi}, \bibinfo{person}{Li-yi Wei}, \bibinfo{person}{Stephen DiVerdi}, \bibinfo{person}{Wilmot Li}, {and} \bibinfo{person}{Daniel Leithinger}.} \bibinfo{year}{2020}\natexlab{}.
\newblock \showarticletitle{RealitySketch: Embedding Responsive Graphics and Visualizations in AR through Dynamic Sketching}. In \bibinfo{booktitle}{\emph{Proceedings of the 33rd Annual ACM Symposium on User Interface Software and Technology}} (Virtual Event, USA) \emph{(\bibinfo{series}{UIST '20})}. \bibinfo{publisher}{Association for Computing Machinery}, \bibinfo{address}{New York, NY, USA}, \bibinfo{pages}{166–181}.
\newblock
\showISBNx{9781450375146}
\urldef\tempurl%
\url{https://doi.org/10.1145/3379337.3415892}
\showDOI{\tempurl}


\bibitem[Tahara et~al\mbox{.}(2020)]%
        {tahara_retargetable_2020}
\bibfield{author}{\bibinfo{person}{Tomu Tahara}, \bibinfo{person}{Takashi Seno}, \bibinfo{person}{Gaku Narita}, {and} \bibinfo{person}{Tomoya Ishikawa}.} \bibinfo{year}{2020}\natexlab{}.
\newblock \showarticletitle{Retargetable {AR}: {Context}-aware {Augmented} {Reality} in {Indoor} {Scenes} based on {3D} {Scene} {Graph}}. In \bibinfo{booktitle}{\emph{2020 {IEEE} {International} {Symposium} on {Mixed} and {Augmented} {Reality} {Adjunct} ({ISMAR}-{Adjunct})}}. \bibinfo{pages}{249--255}.
\newblock
\urldef\tempurl%
\url{https://doi.org/10.1109/ISMAR-Adjunct51615.2020.00072}
\showDOI{\tempurl}


\bibitem[Technologies(2023)]%
        {unity-website}
\bibfield{author}{\bibinfo{person}{Unity Technologies}.} \bibinfo{year}{2023}\natexlab{}.
\newblock \bibinfo{booktitle}{\emph{Unity}}.
\newblock
\urldef\tempurl%
\url{https://unity.com/}
\showURL{%
\tempurl}
\newblock
\shownote{Accessed on Oct 9th, 2023}.


\bibitem[{Unity Technologies}(2023)]%
        {UnityMARS2023}
\bibfield{author}{\bibinfo{person}{{Unity Technologies}}.} \bibinfo{year}{2023}\natexlab{}.
\newblock \bibinfo{title}{Getting Started with Unity MARS}.
\newblock \bibinfo{howpublished}{\url{https://unity.com/products/mars/get-started}}.
\newblock
\newblock
\shownote{Accessed: 2024-04-02}.


\bibitem[Upchurch and Niu(2022)]%
        {upchurch_dense_2022}
\bibfield{author}{\bibinfo{person}{Paul Upchurch} {and} \bibinfo{person}{Ransen Niu}.} \bibinfo{year}{2022}\natexlab{}.
\newblock \bibinfo{title}{A {Dense} {Material} {Segmentation} {Dataset} for {Indoor} and {Outdoor} {Scene} {Parsing}}.
\newblock
\newblock
\urldef\tempurl%
\url{http://arxiv.org/abs/2207.10614}
\showURL{%
\tempurl}
\newblock
\shownote{arXiv:2207.10614 [cs]}.


\bibitem[Wu et~al\mbox{.}(2023b)]%
        {wu2023visual}
\bibfield{author}{\bibinfo{person}{Chenfei Wu}, \bibinfo{person}{Shengming Yin}, \bibinfo{person}{Weizhen Qi}, \bibinfo{person}{Xiaodong Wang}, \bibinfo{person}{Zecheng Tang}, {and} \bibinfo{person}{Nan Duan}.} \bibinfo{year}{2023}\natexlab{b}.
\newblock \showarticletitle{Visual chatgpt: Talking, drawing and editing with visual foundation models}.
\newblock \bibinfo{journal}{\emph{arXiv preprint arXiv:2303.04671}} (\bibinfo{year}{2023}).
\newblock


\bibitem[Wu et~al\mbox{.}(2023a)]%
        {wu2023large}
\bibfield{author}{\bibinfo{person}{Yusong Wu}, \bibinfo{person}{Ke Chen}, \bibinfo{person}{Tianyu Zhang}, \bibinfo{person}{Yuchen Hui}, \bibinfo{person}{Taylor Berg-Kirkpatrick}, {and} \bibinfo{person}{Shlomo Dubnov}.} \bibinfo{year}{2023}\natexlab{a}.
\newblock \showarticletitle{Large-scale contrastive language-audio pretraining with feature fusion and keyword-to-caption augmentation}. In \bibinfo{booktitle}{\emph{ICASSP 2023-2023 IEEE International Conference on Acoustics, Speech and Signal Processing (ICASSP)}}. IEEE, \bibinfo{pages}{1--5}.
\newblock


\bibitem[Yang et~al\mbox{.}(2023)]%
        {yang2023diffsound}
\bibfield{author}{\bibinfo{person}{Dongchao Yang}, \bibinfo{person}{Jianwei Yu}, \bibinfo{person}{Helin Wang}, \bibinfo{person}{Wen Wang}, \bibinfo{person}{Chao Weng}, \bibinfo{person}{Yuexian Zou}, {and} \bibinfo{person}{Dong Yu}.} \bibinfo{year}{2023}\natexlab{}.
\newblock \showarticletitle{Diffsound: Discrete diffusion model for text-to-sound generation}.
\newblock \bibinfo{journal}{\emph{IEEE/ACM Transactions on Audio, Speech, and Language Processing}} (\bibinfo{year}{2023}).
\newblock


\bibitem[Zhou et~al\mbox{.}(2018)]%
        {Zhou_2018_CVPR}
\bibfield{author}{\bibinfo{person}{Yipin Zhou}, \bibinfo{person}{Zhaowen Wang}, \bibinfo{person}{Chen Fang}, \bibinfo{person}{Trung Bui}, {and} \bibinfo{person}{Tamara~L. Berg}.} \bibinfo{year}{2018}\natexlab{}.
\newblock \showarticletitle{Visual to Sound: Generating Natural Sound for Videos in the Wild}. In \bibinfo{booktitle}{\emph{Proceedings of the IEEE Conference on Computer Vision and Pattern Recognition (CVPR)}}.
\newblock


\bibitem[Zhou et~al\mbox{.}(2007)]%
        {zhou_role_2007}
\bibfield{author}{\bibinfo{person}{ZhiYing Zhou}, \bibinfo{person}{Adrian~David Cheok}, \bibinfo{person}{Yan Qiu}, {and} \bibinfo{person}{Xubo Yang}.} \bibinfo{year}{2007}\natexlab{}.
\newblock \showarticletitle{The {Role} of 3-{D} {Sound} in {Human} {Reaction} and {Performance} in {Augmented} {Reality} {Environments}}.
\newblock \bibinfo{journal}{\emph{IEEE Transactions on Systems, Man, and Cybernetics - Part A: Systems and Humans}} \bibinfo{volume}{37}, \bibinfo{number}{2} (\bibinfo{date}{March} \bibinfo{year}{2007}), \bibinfo{pages}{262--272}.
\newblock
\showISSN{1558-2426}
\urldef\tempurl%
\url{https://doi.org/10.1109/TSMCA.2006.886376}
\showDOI{\tempurl}
\newblock
\shownote{Conference Name: IEEE Transactions on Systems, Man, and Cybernetics - Part A: Systems and Humans}.


\end{thebibliography}

\appendix

\newpage
\onecolumn
\rev{
\section{Task Prompts}
\label{appendix}
Here we provide text prompts for tuning a GPT4.0 model into our sound acquisition process controller.
\newline
\newline
\fbox{
  \parbox{\textwidth}{
\textbf{System Context:}

You are an assistant helping user generate or retrieve matching sound assets for augmented reality events. You will be provided with the description of the event, including: who caused this event, what is this event, and what is the subject of this event, which can be null for some cases. Note that the description of the causer and subject of event can sometimes be long and you need to extract the important part of it to create concise replies. For example, when being told that the causing entity of the event is ``metalball of a plastic slope with balls running on it'' you should know that the only mattering keyword here is the metalball.

You have three methods to provide sound assets. Method 1 is recommending from a predefined list of sound assets, each with a name describing the content. You will recommend the best matching name based on event description. The full list of sound asset will be provided at the end of this context; Method 2 is an online sound retrieval API, for which you should generate a simplified text prompt to search with, preferably within 4 words; Method 3 is a diffusion model that takes text prompt and generate sound assets. This method has three functions: generating new sound file with text prompt, generating similar sound file based on an existing sound, and transferring an existing sound with a text prompt.

For each conversation, you will first provide a response about all three sound sources. User may ask following up questions in reply, then you just reply based on specific feedback.
When providing results, you should always follow a strict format starting and end with \#. For method 1, you should reply the best matching sound asset in \#method1:FILENAME\# format. For method 2, you should reply in format of \#method2:PROMPT\#, please replace the PROMPT with your generated prompt. For method 3, when generating new sound, you should reply with \#method3generation:PROMPT\#, please replace the PROMPT with your generated prompt; when generating similar sound of a existing sound, reply with \#method3similar\#, when transferring an existing sound, reply with \#method3transfer:PROMPT\#, please replace the PROMPT with your generated prompt. Note that the hashtags need to be kept at both start and end of the result, the FILENAME need to be replaced by the exact name provided in the asset list, with no exception. The PROMPT should be replaced with the prompt you generate for these cases. Don't explain why you have these prompts, just stricktly follow the formats.

Here I list names of sound assets to recommend from, each filename represents its content: 

1. Knock surface

2. Sliding sound

3. Crash Bulb Break

\textit{<... 32 assets in total...>}
}
}
\newline
\newline
We also prompt GPT4.0 to better parse the context information of specific AR events. We show one example event in the following prompt:
\newline
\newline
\fbox{
  \parbox{\textwidth}{
\textbf{Conversation Prompt:}

Please give me the results as hinted by context. For this specific event, the description of the AR event is:

The sound-producing AR event is \textit{TapSurface}. The causing entity of this event is \textit{a metal robot}. The target entity is \textit{a wooden surface}.

You should reply top5 results with method 1, ensuring that the filename provided is exactly from the asset list. You should also give one result with method 3 transfer, with a reasonable prompt that could transfer an existing tapping sound to a more material-aware sound. For example, for a tapping on a wooden surface, you could use tapping wood as the prompt. Reply with the format of \#method3transfer:PROMPT\#, replacing the PROMPT with your generated prompt. You can add a 'high quality' tag in the prompt to help improve the sound quality.
}}
}

\end{document}